\documentclass[aps,showpacs,preprintnumbers,amsmath,amssymb,twocolumn,superscriptaddress,prx,floatfix,nofootbib]{revtex4-1}

\usepackage{amsfonts}
\usepackage{amsmath}
\usepackage{hyperref}
\usepackage{amssymb}
\usepackage{graphicx}
\usepackage{dcolumn}
\usepackage{bm}
\usepackage{xkcdcolors}
\usepackage{tabularx}
\usepackage{epstopdf}
\usepackage{xcolor}
\usepackage{mathrsfs}
\usepackage{subfig}
\usepackage{soul}
\usepackage{mathtools}
\usepackage{float}
\usepackage{verbatim}
\usepackage{comment}
\usepackage{cleveref}
\usepackage{ragged2e}
\usepackage{caption}
\DeclareCaptionJustification{justified}{\justifying}
\begin{document}

\title{Deterministic, quenched and annealed parameter estimation\\for heterogeneous network models}

\author{Marzio Di Vece}
\email{marzio.divece@imtlucca.it}
\affiliation{IMT School for Advanced Studies, Piazza San Francesco 19, 55100 Lucca, Italy}
\affiliation{Scuola Normale Superiore, P.zza dei Cavalieri 7, 56126 Pisa, Italy}
\author{Diego Garlaschelli}
\affiliation{IMT School for Advanced Studies, Piazza San Francesco 19, 55100 Lucca, Italy}
\affiliation{Lorentz Institute for Theoretical Physics, University of Leiden, Niels Bohrweg 2, 2333 CA Leiden, The Netherlands}
\affiliation{INdAM-GNAMPA Istituto Nazionale di Alta Matematica, 00185 Rome, Italy}
\author{Tiziano Squartini}
\affiliation{IMT School for Advanced Studies, Piazza San Francesco 19, 55100 Lucca, Italy}
\affiliation{Scuola Normale Superiore, P.zza dei Cavalieri 7, 56126 Pisa, Italy}
\affiliation{INdAM-GNAMPA Istituto Nazionale di Alta Matematica, 00185 Rome, Italy}
\affiliation{Institute for Advanced Study, University of Amsterdam, Oude Turfmarkt 145, 1012 GC Amsterdam, The Netherlands}

\date{\today}

\begin{abstract}
At least two, different approaches to define and solve statistical models for the analysis of economic systems exist: the typical, econometric one, interpreting the Gravity Model specification as the expected link weight of an arbitrary probability distribution, and the one rooted into statistical physics, constructing maximum-entropy distributions constrained to satisfy certain network properties. In a couple of recent, companion papers they have been successfully integrated within the framework induced by the constrained minimisation of the Kullback-Leibler divergence: specifically, two, broad classes of models have been devised, i.e. the integrated and the conditional ones, defined by different, probabilistic rules to place links, load them with weights and turn them into proper, econometric prescriptions. Still, the recipes adopted by the two approaches to estimate the parameters entering into the definition of each model differ. In econometrics, a likelihood that decouples the binary and weighted parts of a model, treating a network as deterministic, is typically maximised; to restore its random character, two alternatives exist: either solving the likelihood maximisation on each configuration of the ensemble and taking the average of the parameters afterwards or taking the average of the likelihood function and maximising the latter one. The difference between these approaches lies in the order in which the operations of `averaging' and `maximisation' are taken - a difference that is reminiscent of the `quenched’ and `annealed’ ways of averaging out the disorder in spin glasses. The results of the present contribution, devoted to comparing these recipes in the case of continuous, conditional network models, indicate that the `annealed' estimation recipe represents the best alternative to the `deterministic' one.
\end{abstract}


\maketitle

\section{Introduction}

Over the last twenty years, the growth of network science has impacted several disciplines by establishing new, empirical facts about the structural properties of the related systems. Prominent examples are provided by economics and finance: the growing availability of data has motivated researchers to explore and model the architecture of cryptocurrencies~\cite{Vallarano2020}, interbank networks~\cite{Bardoscia2021}, production networks~\cite{Ialongo2022} and trading networks~\cite{Garlaschelli2004,Schweitzer2009,Fronczak2012b,Herman2022}.

Modelling the establishment of a connection and the corresponding weight simultaneously poses a serious challenge. Econometrics prescribes to estimate binary and weighted parameters either separately, within the context of hurdle models~\cite{Mullahy}, or jointly, within the context of zero-inflated models~\cite{Burger2009}; in both cases, the Gravity Model specification~\cite{Tinbergen1962} $\langle w_{ij}\rangle_\text{GM}=f(\omega_i,\omega_j,d_{ij}|\underline{\phi})=e^{\rho}(\omega_i\omega_j)^{\alpha}d_{ij}^{\gamma}$ - where $\omega_i\equiv{\text{GDP}_i}/{\overline{\text{GDP}}}$ is the GDP of country $i$ divided by the arithmetic mean of the GDPs of all countries, $d_{ij}$ is the geographic distance between the capitals of countries $i$ and $j$ and $\underline{\phi}\equiv(\rho,\alpha,\gamma)$ is the vector of parameters defining the Gravity Model specification - is interpreted as the expected value of a probability distribution whose functional form is arbitrary. On the other hand, the approach rooted in statistical physics constructs maximum-entropy distributions, constrained to satisfy certain network properties~\cite{Shannon,Jaynes1957a,Cover,DSBook,Cimini2019}. 

In a couple of recent, companion papers~\cite{Marzio2022,Marzio2023} the two, aforementioned approaches have been integrated within the framework induced by the constrained optimisation of the Kullback-Leibler (KL) divergence~\cite{Kullback1951}. In particular, two, broad classes of models have been constructed, i.e. the integrated and conditional ones, defined by different, probabilistic rules to place links, load them with weights and turn them into properly econometric prescriptions. For what concerns integrated models, the first, two rules follow from a single, constrained optimisation of the KL divergence~\cite{Garlaschelli2009}; for what concerns conditional models, the two rules are disentangled and the functional form of the weight distribution follows from a conditional, optimisation procedure~\cite{Parisi2020}. Still, the prescriptions adopted by the two approaches to carry out the estimation of the parameters entering into the definition of each model differ.

The present contribution is devoted to comparing these recipes in the case of continuous, conditional network models defined by both homogeneous and heterogeneous constraints.

\section{Minimisation of the\\Kullback-Leibler divergence}

The functional form of continuous, conditional network models can be identified through the constrained minimisation of the KL divergence of a distribution $Q$ from a prior distribution $R$, i.e.

\begin{equation}
D_\text{KL}(Q||R)=\int_\mathbb{W}Q(\mathbf{W})\ln \frac{Q(\mathbf{W})}{R(\mathbf{W})}d\mathbf{W}
\end{equation}
where $\mathbf{W}$ is one of the possible values of a continuous random variable, $\mathbb{W}$ is the set of possible values that $\mathbf{W}$ can take, $Q(\mathbf{W})$ is the (multivariate) probability density function to be estimated and $R(\mathbf{W})$ plays the role of prior  distribution, whose divergence from $Q(\mathbf{W})$ must be minimised: in our setting, $\mathbf{W}$ represents an entire network whose weights, now, obey the property $w_{ij}\in\mathbb{R}^+_0$, $\forall\:i<j$. Such an optimisation scheme embodies the so-called Minimum Discrimination Information Principle~\cite{Marzio2022,Marzio2023}, implementing the idea that, as new information becomes available, an updated distribution $Q(\mathbf{W})$ should be chosen in order to make its discrimination from the prior distribution $R(\mathbf{W})$ as hard as possible. 

Let us, now, separate both the prior and the posterior distribution into a purely binary part and a conditional, weighted one; the positions $Q(\mathbf{W})=P(\mathbf{A})Q(\mathbf{W}|\mathbf{A})$ and $R(\mathbf{W})=T(\mathbf{A})R(\mathbf{W}|\mathbf{A})$, where $\mathbf{A}$ denotes the binary projection of the weighted network $\mathbf{W}$ (i.e. $\Theta[\mathbf{W}]=\mathbf{A}$), $T(\mathbf{A})$ represents the binary prior and $R(\mathbf{W}|\mathbf{A})$ represents the conditional, weighted prior, lead the KL divergence to be re-writable as 

\begin{equation}
D_\text{KL}(Q||R)=D_\text{KL}(P||T)+D_\text{KL}(\overline{Q}||\overline{R}),
\end{equation}
i.e. as a sum of the two addenda

\begin{align}
D_\text{KL}(P||T)&=\sum_{\mathbf{A}\in\mathbb{A}}P(\mathbf{A})\ln\frac{P(\mathbf{A})}{T(\mathbf{A})},\\
D_\text{KL}(\overline{Q}||\overline{R})&=\sum_{\mathbf{A}\in\mathbb{A}}P(\mathbf{A})\int_{\mathbb{W}_\mathbf{A}}Q(\mathbf{W}|\mathbf{A})\ln\frac{Q(\mathbf{W}|\mathbf{A})}{R(\mathbf{W}|\mathbf{A})}d\mathbf{W}.
\end{align}

In what follows, we will deal with completely uninformative priors, a choice that amounts at considering the (somehow, simplified) expression

\begin{equation}
-S(Q)=-S(P)-S(\overline{Q}|P)
\end{equation}
i.e. `minus' the joint entropy, where

\begin{equation}
S(P)=-\sum_{\mathbf{A}\in\mathbb{A}}P(\mathbf{A})\ln P(\mathbf{A})
\end{equation}
is the Shannon entropy of the probability distribution describing the binary projection of the network structure~\cite{DSBook,Cimini2019} and

\begin{equation}
S(\overline{Q}|P)=-\sum_{\mathbf{A}\in\mathbb{A}}P(\mathbf{A})\int_{\mathbb{W}_\mathbf{A}}Q(\mathbf{W}|\mathbf{A})\ln Q(\mathbf{W}|\mathbf{A})d\mathbf{W}
\end{equation}
is the conditional Shannon entropy of the probability distribution describing the weighted network structure~\cite{Marzio2022,Marzio2023,Parisi2020}. Notice that, when continuous models are considered, $S(\overline{Q}|P)$ is defined by a sum running over all the binary configurations within the ensemble $\mathbb{A}$ and an integral over all the weighted configurations that are compatible with each, specific, binary structure, i.e. $\mathbb{W}_\mathbf{A}=\{\mathbf{W}:\Theta[\mathbf{W}]=\mathbf{A}\}$. For a more detailed discussion, see Appendix A.

The functional form of $P(\mathbf{A})$ can be determined by carrying out the usual, constrained maximisation of Shannon entropy~\cite{DSBook,Cimini2019}; remarkably, any set of (binary) constraints considered in the present paper will lead to the same expression for $P(\mathbf{A})$, i.e. $P(\mathbf{A})=\prod_{i<j}p_{ij}^{a_{ij}}(1-p_{ij})^{1-a_{ij}}$ with $p_{ij}=x_{ij}/(1+x_{ij})$: specifically, the position $x_{ij}\equiv x$ individuates the Undirected Binary Random Graph Model (UBRGM), the position $x_{ij}\equiv x_ix_j$ individuates the Undirected Binary Configuration Model (UBCM) and the position $x_{ij}\equiv \delta\omega_i\omega_j$ individuates the Logit Model (LM)~\cite{Caldarelli2003}.

On the other hand, the functional form of $Q(\mathbf{W}|\mathbf{A})$ can be determined by carrying out the constrained maximisation of $S(\overline{Q}|P)$, the set of constraints being, now,

\begin{align}
1&=\int_{\mathbb{W}_\mathbf{A}}P(\mathbf{W}|\mathbf{A})d\mathbf{W},\:\forall\:\mathbf{A}\in\mathbb{A},\\
\langle C_\alpha\rangle&=\sum_{\mathbf{A}\in\mathbb{A}}P(\mathbf{A})\int_{\mathbb{W}_\mathbf{A}}Q(\mathbf{W}|\mathbf{A})C_\alpha(\mathbf{W})d\mathbf{W},\:\forall\:\alpha;
\end{align}
while the first condition ensures the normalisation of the probability distribution, the vector $\{C_\alpha(\mathbf{W})\}$ represents the proper set of weighted constraints. The distribution induced by such an optimisation problem reads

\begin{equation}
Q(\mathbf{W}|\mathbf{A})=\frac{e^{-H(\mathbf{W})}}{Z_\mathbf{A}}=\frac{e^{-H(\mathbf{W})}}{\int_{\mathbb{W}_\mathbf{A}}e^{-H(\mathbf{W})}d\mathbf{W}} 
\end{equation}
if $\mathbf{W}\in\mathbb{W}_\mathbf{A}$ and $0$ otherwise. While the Hamiltonian $H(\mathbf{W})=\sum_\alpha\psi_\alpha C_\alpha(\mathbf{W})$ lists the constraints, the quantity at the denominator is the partition function, conditional on the fixed topology $\mathbf{A}$~\cite{Parisi2020}. 

For mathematical convenience, in what follows we will consider separable Hamiltonians, i.e. functions that can be written as sums of node pair-specific Hamiltonians: $H(\mathbf{W})=\sum_{i<j}H_{ij}(w_{ij})$; this choice leads to the result

\begin{align}\label{eq_19}
Q(\mathbf{W}|\mathbf{A})&=\frac{e^{-\sum_{i<j}H_{ij}(w_{ij})}}{\int_{\mathbb{W}_\mathbf{A}}e^{-\sum_{i<j}H_{ij}(w_{ij})}d\mathbf{W}}\nonumber\\
&=\prod_{i<j}\frac{e^{-H_{ij}(w_{ij})}}{\left[\int_{m_{ij}}^{+\infty}e^{-H_{ij}(w_{ij})}dw_{ij}\right]^{a_{ij}}}=\prod_{i<j}\frac{e^{-H_{ij}(w_{ij})}}{\zeta_{ij}^{a_{ij}}}
\end{align}
(with $m_{ij}$ being the pair-specific, minimum weight allowed by a given model and $\zeta_{ij}$ being the corresponding partition function), irrespectively from the specific, functional form of $H_{ij}(w_{ij})$~\cite{Marzio2023}. For a more detailed discussion, see Appendix B.

\section{Estimation of the parameters}

Several, alternative recipes are viable to estimate the parameters entering into the definition of continuous, conditional network models.

\subsection{`Deterministic' parameter estimation}

The simplest one prescribes to consider the traditional likelihood function

\begin{align}
\ln Q(\mathbf{W}^*)&=\ln[P(\mathbf{A}^*)Q(\mathbf{W}^*|\mathbf{A}^*)]\nonumber\\
&=\ln P(\mathbf{A}^*)+\ln Q(\mathbf{W}^*|\mathbf{A}^*)
\end{align}
with $\mathbf{W}^*$ ($\mathbf{A}^*$) being the empirical, weighted (binary) adjacency matrix; its maximisation allows the parameters entering into the definition of the purely topological distribution and those entering into the definition of the conditional, weighted one to be estimated in a totally disentangled fashion~\cite{Marzio2023}. In fact, maximising

\begin{align}
\mathcal{L}_{\underline{\psi}}&=\ln Q(\mathbf{W}^*|\mathbf{A}^*)\nonumber\\
&=-H(\mathbf{W}^*)-\ln Z_{\mathbf{A}^*}\nonumber\\
&=-H(\mathbf{W}^*)-\ln\left[\int_{\mathbb{W}_{\mathbf{A}^*}}e^{-H(\mathbf{W})}d\mathbf{W}\right]
\end{align}
with respect to the unknown parameters leads us to find the vector of values $\underline{\psi}^*$ satisfying the vector of relationships

\begin{equation}
\langle\mathbf{C}\rangle_{\mathbf{A}^*}(\underline{\psi}^*)\equiv\mathbf{C}^*
\end{equation}
which stands for the set of relationships $\langle C_\alpha\rangle_{\mathbf{A}^*}(\underline{\psi}^*)\equiv C_\alpha^*$, $\forall\:\alpha$, each one equating the model-induced, average value of the corresponding constraint to its empirical value, marked with an asterisk.

This first approach to parameter estimation can be named as `deterministic', to stress that $\mathbf{A}^*$ is considered as not being subject to variation; otherwise stated, this recipe - which is the most common in econometrics - prescribes to estimate the parameters entering into the definition of the conditional, weighted probability distribution by assuming the network topology to be fixed.

\subsection{`Annealed' parameter estimation}

Topology, however, is a random variable itself, obeying the probability distribution $P(\mathbf{A})$. As a consequence, the `deterministic' recipe for parameter estimation could lead to inconsistencies, should the description of $\mathbf{A}^*$ provided by $P(\mathbf{A})$ be not accurate. The variability induced by $P(\mathbf{A})$ can be properly accounted for by considering the generalised likelihood~\cite{Parisi2020}

\begin{align}
\mathcal{G}_{\underline{\psi}}&=\sum_{\mathbf{A}\in\mathbb{A}}P(\mathbf{A})\ln Q(\mathbf{W}^*|\mathbf{A})\nonumber\\
&=\sum_{\mathbf{A}\in\mathbb{A}}P(\mathbf{A})[-H(\mathbf{W}^*)-\ln Z_\mathbf{A}]\nonumber\\
&=-H(\mathbf{W}^*)-\sum_{\mathbf{A}\in\mathbb{A}}P(\mathbf{A})\ln\left[\int_{\mathbb{W}_{\mathbf{A}^*}}e^{-H(\mathbf{W})}d\mathbf{W}\right]=\langle\mathcal{L}_{\underline{\psi}}\rangle
\end{align}
whose maximisation leads us to find the vector of values $\underline{\psi}^*$ satisfying the vector of relationships

\begin{equation}
\sum_{\mathbf{A}\in\mathbb{A}}P(\mathbf{A})\langle\mathbf{C}\rangle_{\mathbf{A}}(\underline{\psi}^*)=\langle\mathbf{C}\rangle(\underline{\psi}^*)=\mathbf{C^*}
\end{equation}
which stands for the set of relationships $\langle C_\alpha\rangle(\underline{\psi}^*)\equiv C_\alpha^*$, $\forall\:\alpha$. Taking this average is conceptually similar to taking the `annealed' average in physics: parameter estimation is carried out while random variables - again, the entries of the adjacency matrix - are left to vary.

Interestingly, the `deterministic' recipe is a special case of the `annealed' recipe since the former can be recovered by posing $P(\mathbf{A})\equiv\delta_{\mathbf{A},\mathbf{A}^*}$: in this case, in fact,

\begin{align}
\mathcal{G}_{\underline{\psi}}&=-H(\mathbf{W}^*)-\sum_{\mathbf{A}\in\mathbb{A}}\delta_{\mathbf{A},\mathbf{A}^*}\ln Z_{\mathbf{A}}\nonumber\\
&=-H(\mathbf{W}^*)-\ln Z_{\mathbf{A}^*}=\mathcal{L}_{\underline{\psi}};
\end{align}
similarly, $\sum_{\mathbf{A}\in\mathbb{A}}\delta_{\mathbf{A},\mathbf{A}^*}\langle\mathbf{C}\rangle_{\mathbf{A}}(\underline{\psi}^*)=\langle\mathbf{C}\rangle_{\mathbf{A}^*}(\underline{\psi}^*)=\mathbf{C^*}$.

\subsection{`Quenched' parameter estimation}

A viable alternative to properly account for the variability induced by $P(\mathbf{A})$ is that of reversing the two operations of `likelihood maximisation' and `ensemble averaging': in other words, one can 1) numerically sample the ensemble of configurations induced by $P(\mathbf{A})$, 2) maximise the likelihood $\ln Q(\mathbf{W}^*|\mathbf{A})$ for each, generated network, 3) take the average of the resulting set of parameters, according to the formula

\begin{equation}
\sum_{\mathbf{A}\in\mathbb{A}}P(\mathbf{A})\underline{\psi}^*(\mathbf{A})=\langle\underline{\psi}^*\rangle
\end{equation}
the estimation of the $\alpha$-th parameter being assumed to coincide with the average $\langle\psi_\alpha^*\rangle$.

Taking this average is conceptually similar to taking the `quenched' average in physics: random variables - in the specific case, the entries of the adjacency matrix - are frozen, parameter estimation is carried out and, only at the end, the values of the parameters are averaged over the ensemble of configurations induced by $P(\mathbf{A})$.\\

As our models inherit their functional form from the constrained minimisation of the KL divergence, each parameter controls for a specific constraint: when employing the `deterministic' recipe, such a circumstance makes each parameter configuration-dependent; when employing either the `annealed' or the `quenched' recipe, instead, accounting for the variability of a network structure induces a sort of `loss of memory' about its empirical, purely topological details.

\section{Results}

In order to test if the `deterministic', `annealed' and `quenched' prescriptions lead to the same estimation, let us focus on a number of variants of the Conditional Exponential Model (CEM), induced by the positions $H_{ij}^\text{CEM}=\beta_{ij}w_{ij}$ and $\zeta_{ij}^\text{CEM}=\beta_{ij}^{-1}$:

\begin{align}
Q(\mathbf{W})&=P(\mathbf{A})Q(\mathbf{W}|\mathbf{A})\nonumber\\
&=\prod_{i<j}p_{ij}^{a_{ij}}(1-p_{ij})^{1-a_{ij}}\prod_{i<j}\beta_{ij}^{a_{ij}}e^{-\beta_{ij}w_{ij}};
\end{align}
naturally, $q_{ij}(w_{ij}=0|a_{ij}=0)=1$ (i.e. if nodes $i$ and $j$ are not connected, the weight of the corresponding link is zero with probability equal to one) and $q_{ij}(w_{ij}>0|a_{ij}=1)=\beta_{ij}e^{-\beta_{ij}w_{ij}}$.

In what follows, we will consider three, different instances of $p_{ij}=x_{ij}/(1+x_{ij})$, corresponding to

\begin{itemize}
\item[$\bullet$] the Undirected Binary Random Graph Model (UBRGM), defined by posing $x_{ij}\equiv x$ and induced by the maximisation of $S(P)$ while constraining the total number of links, $L(\mathbf{A}^*)\equiv L^*=\sum_{i<j}a_{ij}^*$, i.e.

\begin{equation}
p_{ij}^\text{UBRGM}\equiv\frac{x}{1+x};
\end{equation}
\item[$\bullet$] the Undirected Binary Configuration Model (UBCM), defined by posing $x_{ij}\equiv x_ix_j$ and induced by the maximisation of $S(P)$ while constraining the whole degree sequence, $\{k_i(\mathbf{A}^*)\}_{i=1}^N\equiv\{k_i^*\}_{i=1}^N$ with $k_i^*=\sum_{j(\neq i)}a_{ij}^*$, i.e.

\begin{equation}
p_{ij}^\text{UBCM}\equiv\frac{x_ix_j}{1+x_ix_j};
\end{equation}
\item[$\bullet$] two, different instances of the Logit Model (LM), both representing a fitness-driven version of the UBCM, (again) induced by constraining the total number of links, $L(\mathbf{A}^*)\equiv L^*=\sum_{i<j}a_{ij}^*$. The first one is defined by posing $x_{ij}\equiv\delta\omega_i\omega_j$, i.e.

\begin{equation}
p_{ij}^\text{LM}\equiv\frac{\delta\omega_i\omega_j}{1+\delta\omega_i\omega_j}
\end{equation}
and has been employed to study the year 2017 of the CEPII-BACI version of the World Trade Web (WTW)~\cite{Baci2014}, that is a network of $N=171$ nodes and a link density of $d=0.87$. The second one is defined by posing $x_{ij}\equiv\delta s_i s_j$, i.e.

\begin{equation}
p_{ij}^\text{LM}=\frac{\delta s_i s_j}{1+\delta s_i s_j}
\end{equation}
and has been employed to study the 01/03/2019 snapshot of the Bitcoin Lightning Network (BLN)~\cite{BLN}, that is a network of $N=5012$ nodes and a link density of $d=0.003$.
\end{itemize}

\begin{figure}[t!]
\centering
\includegraphics[width=0.49\textwidth]{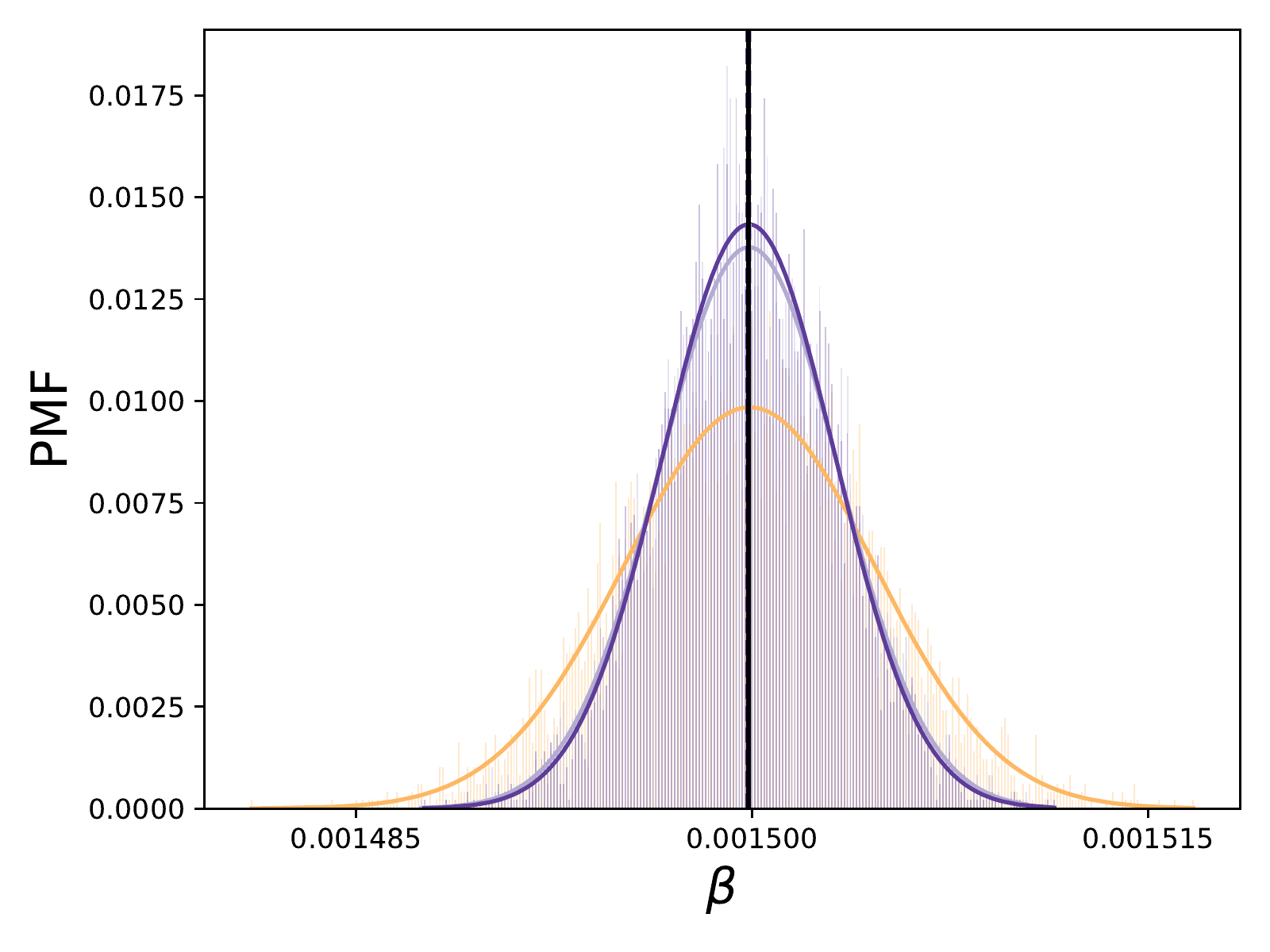}
\caption{Estimations of the parameter $\beta$, entering the definition of the homogeneous version of the CEM, where the binary topology is either `deterministic' (black vertical line) or generated via the UBRGM (light orange or light grey), the UBCM (purple or dark grey) and the LM (light purple or grey). The deterministic approach leads to a single estimate, while the other approaches lead to either a single, `annealed' estimate (vertical, solid lines) or to a whole distribution of `quenched' estimates (empirical distribution constructed over an ensemble of 5.000 binary configurations with theoretical curves, Binomial or Poisson-Binomial, dependent on the binary model; the corresponding average value is indicated by a vertical, dash-dotted line). The `annealed' parameter estimates, the average values of the `quenched' parameter distributions and the `deterministic' parameter estimate coincide. Data refers to the year 2017 of the CEPII-BACI version of the WTW~\cite{Baci2014}.}
\label{fig1}
\end{figure}

\subsection{`Scalar' variant of the\\Conditional Exponential Model}

Let us start by considering the `scalar' or homogeneous variant of the CEM, defined by the position $\beta_{ij}\equiv\beta$, $\forall\:i<j$.

In this case, the `deterministic' recipe for parameter estimation prescribes to maximise the likelihood

\begin{align}
\mathcal{L}_{\underline{\psi}}&=\sum_{i<j}[-\beta w_{ij}^*+a_{ij}^*\ln\beta]=-\beta W^*+L^*\ln\beta
\end{align}
where $W(\mathbf{W}^*)\equiv W^*=\sum_{i<j}w_{ij}^*$ and whose optimisation leads to the expression $\beta=L^*/W^*$. The `annealed' recipe prescribes to maximise the likelihood

\begin{align}
\mathcal{G}_{\underline{\psi}}&=\sum_{i<j}[-\beta w_{ij}^*+p_{ij}\ln\beta]=-\beta W^*+\langle L\rangle\ln\beta
\end{align}
whose optimisation leads to the expression $\beta=\langle L\rangle/W^*$. The `quenched' recipe, on the other hand, prescribes to calculate the average

\begin{equation}
\langle\beta\rangle=\sum_{\mathbf{A}\in\mathbb{A}}P(\mathbf{A})\beta(\mathbf{A})=\sum_{\mathbf{A}\in\mathbb{A}}P(\mathbf{A})\frac{L(\mathbf{A})}{W^*}=\frac{\langle L\rangle}{W^*}
\end{equation}
since, now, $\beta(\mathbf{A})=L(\mathbf{A})/W^*$.

In the case of the `scalar' variant of the CEM, the estimations coincide for any null model preserving the total number of links, i.e. ensuring that $\langle L\rangle=L^*$, regardless of the network density. Such a result is confirmed by Fig.~\ref{fig1} where each recipe has been implemented on the WTW, by adopting the distributions induced by the UBRGM (blue), the UBCM (green) and the LM (red). Specifically, the `deterministic' estimation (black, solid line) and the `annealed' estimations (blue, green and red, solid lines) overlap; moreover, each `annealed' estimation overlaps with the the corresponding, `quenched' estimation, i.e. the average value of the related, `quenched' distribution (blue, green and red, dash-dotted lines).

In the case of the UBRGM-induced, homogeneous version of the CEM, the `quenched' distribution of the parameter $\beta(\mathbf{A})=L(\mathbf{A})/W^*$ `inherits' the distribution of the total number of links, i.e. $L\sim\text{Bin}(N(N-1)/2,p)$, with $p=2L^*/N(N-1)$: more precisely, $W\beta\sim\text{Bin}(N(N-1)/2,p)$; analogously for the UBCM- and the LM-induced, homogeneous versions of the CEM - the only difference being that, now, $L$ obeys two, different, Poisson-Binomial distributions.

\begin{figure}[t!]
\centering
\includegraphics[width=0.49\textwidth]{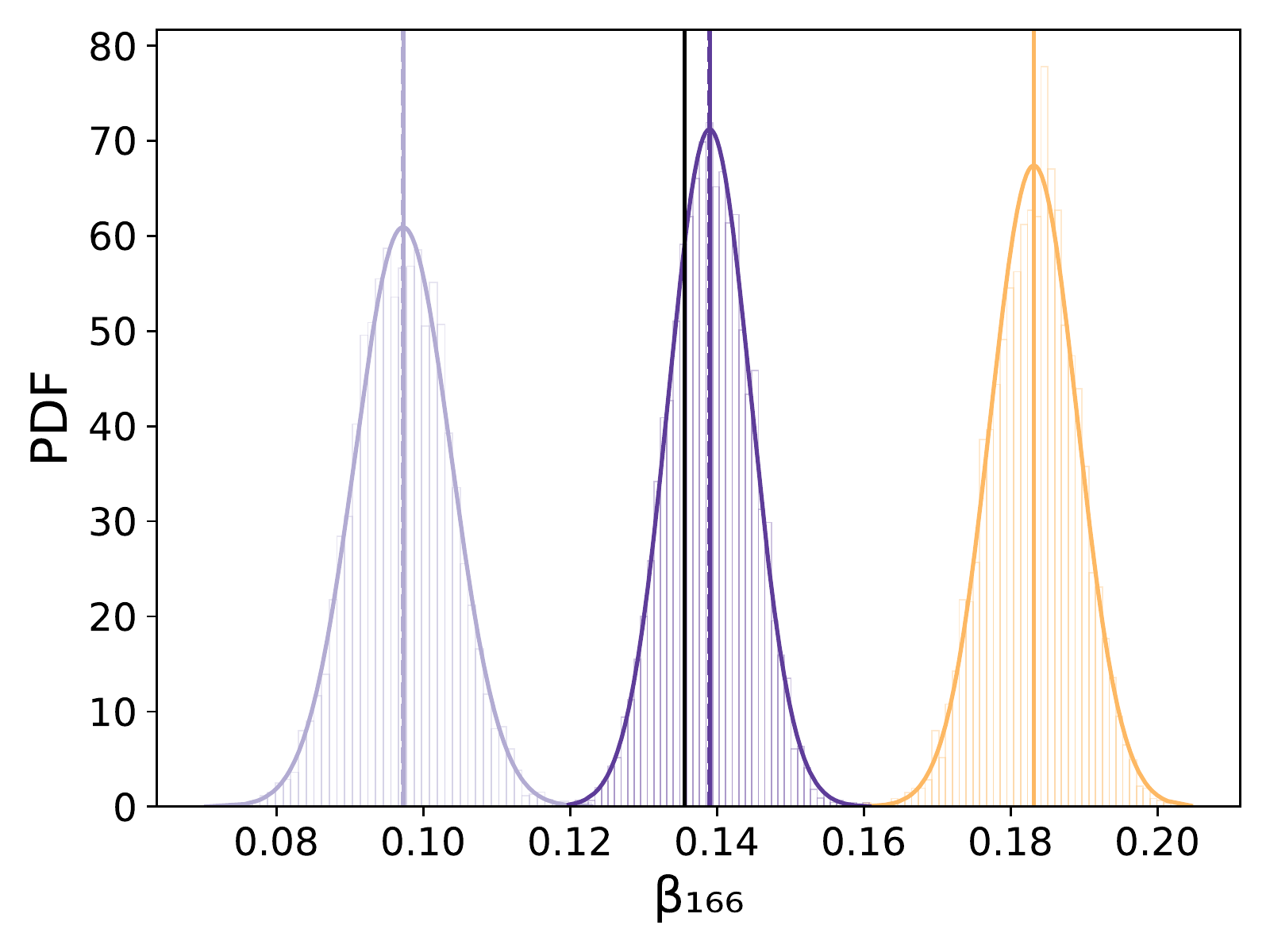}
\caption{Estimations of the parameter $\beta_{166}$ entering the definition of the weakly heterogeneous version of the CEM, where the binary topology is either `deterministic' (black vertical line) or generated via the UBRGM (light orange or light grey), the UBCM (purple or dark grey) and the LM (light purple or grey). The deterministic approach leads to a single estimate, while the other approaches lead to either a single, `annealed' estimate (vertical, solid lines) or to a whole distribution of `quenched' estimates (histograms with normal density curves having the same average and standard deviation, constructed over an ensemble of 5.000 binary configurations; the average value is indicated by a vertical, dash-dotted line). Each `annealed' parameter estimate coincides with the average value of the corresponding `quenched' distribution although the distributions induced by the three, binary recipes are well separated. In addition, the `deterministic' parameter estimate is very close to the UBCM-induced, `annealed' one. Data refers to the year 2017 of the CEPII-BACI version of the WTW~\cite{Baci2014}.}
\label{fig2}
\end{figure}

\subsection{`Vector' variant of the\\Conditional Exponential Model}

Let us, now, consider the `vector' or weakly heterogeneous variant of the CEM, defined by the position $\beta_{ij}\equiv\beta_i+\beta_j$, $\forall\:i<j$.

In this case, the `deterministic' recipe for parameter estimation prescribes to maximise the likelihood

\begin{align}
\mathcal{L}_{\underline{\psi}}&=\sum_{i<j}[-(\beta_i+\beta_j)w_{ij}^*+a_{ij}^*\ln(\beta_i+\beta_j)]\nonumber\\
&=-\sum_i\beta_is_i^*+\sum_{i<j}a_{ij}^*\ln(\beta_i+\beta_j)
\end{align}
where $s_i(\mathbf{W}^*)\equiv s_i^*=\sum_{j(\neq i)}w_{ij}^*$ and whose optimisation requires to solve the system of equations

\begin{equation}
s_i^*=\sum_{j(\neq i)}\frac{a_{ij}^*}{\beta_i+\beta_j},\:\forall\:i.
\end{equation}

\begin{figure*}[t!]
\centering
\includegraphics[width=0.49\textwidth]{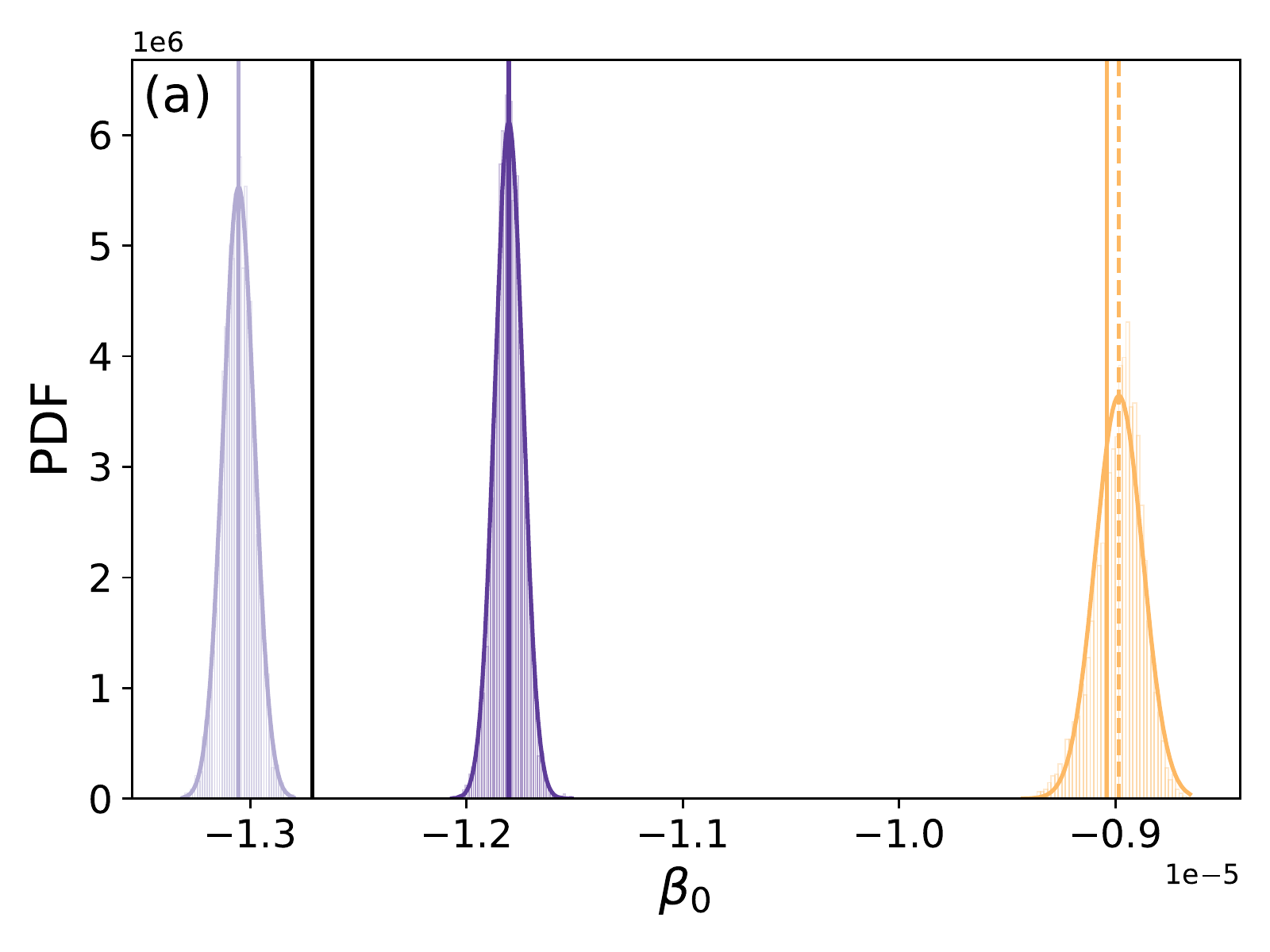}
\includegraphics[width=0.49\textwidth]{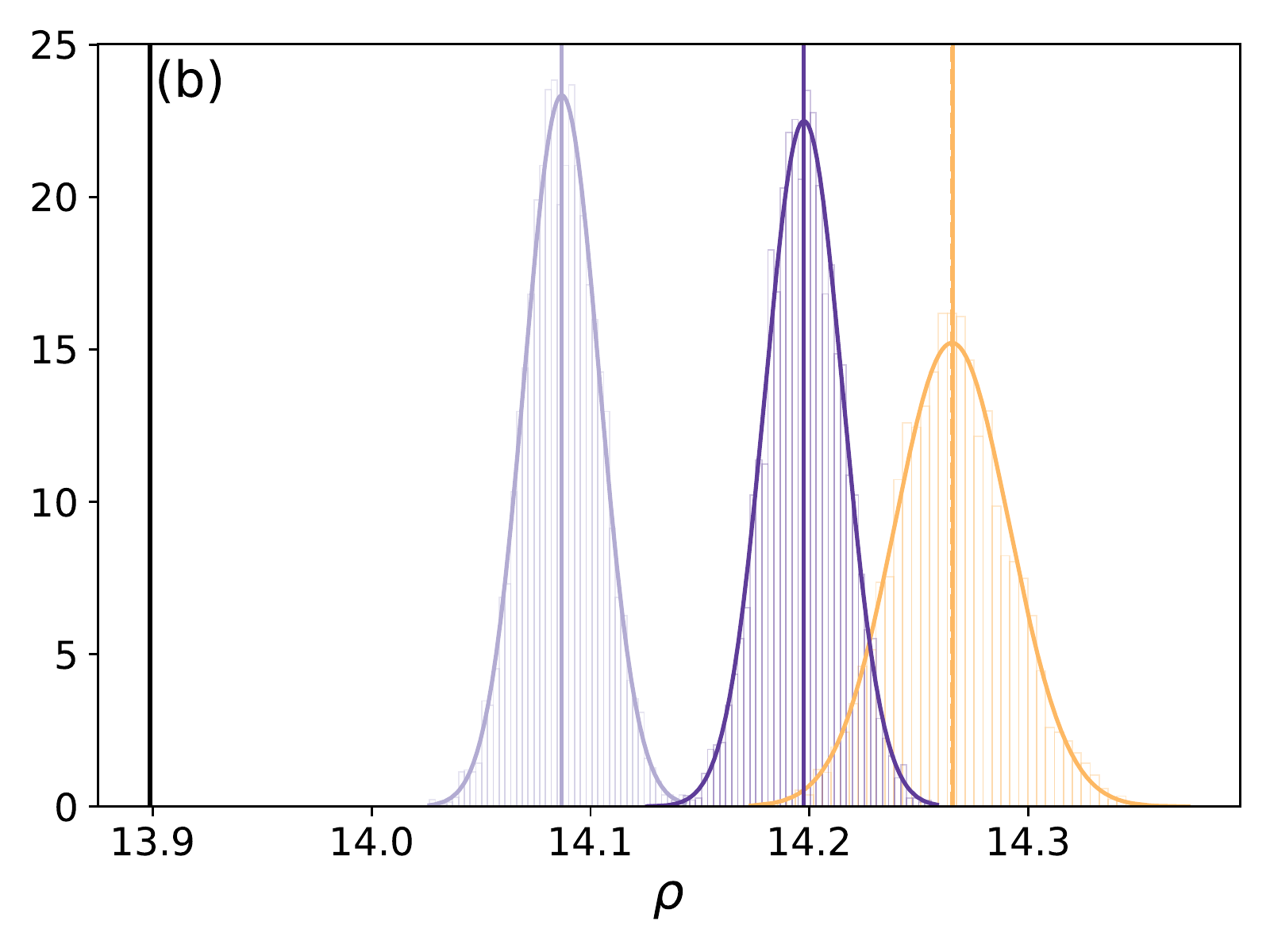}\\
\includegraphics[width=0.49\textwidth]{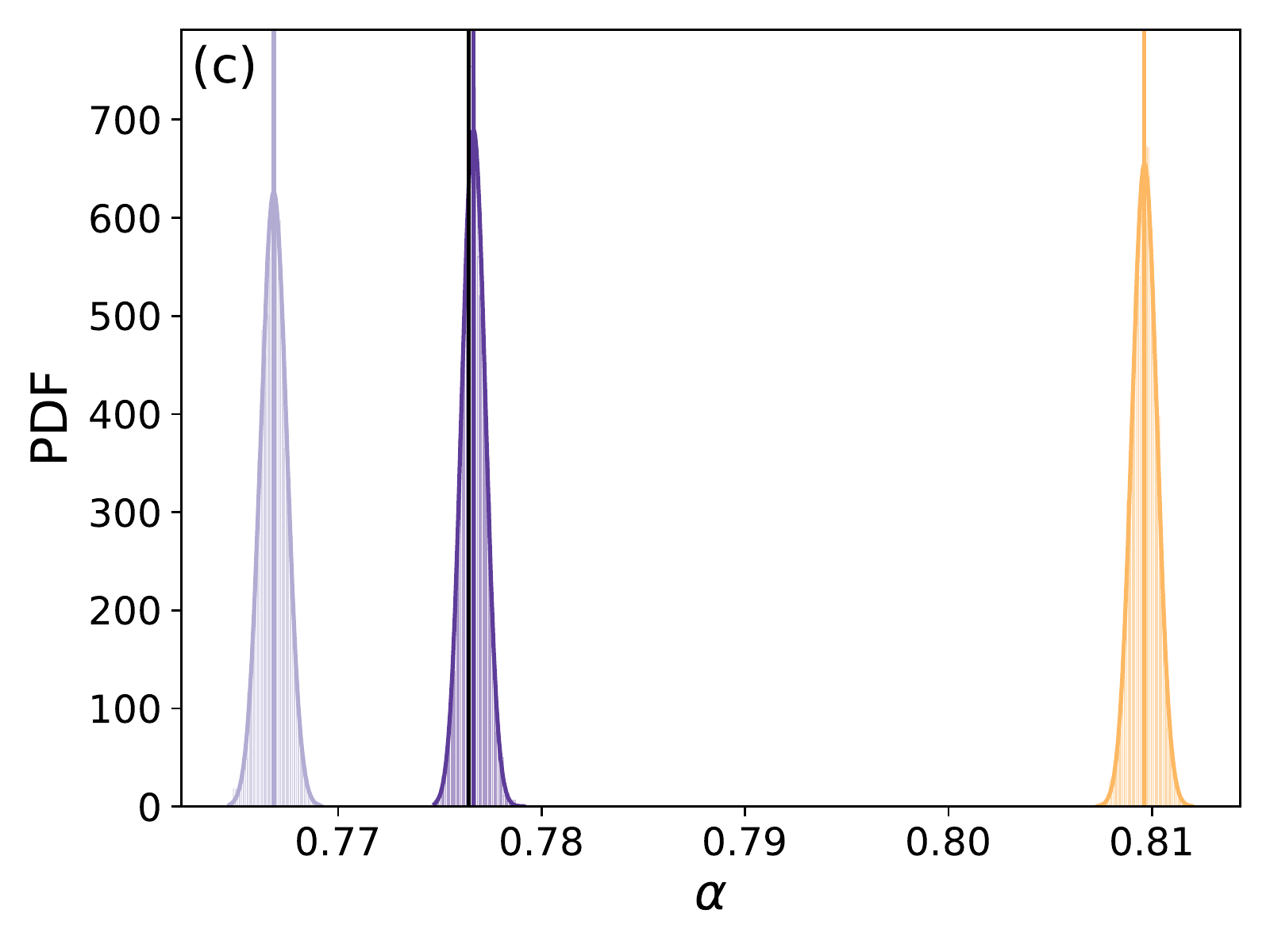}
\includegraphics[width=0.49\textwidth]{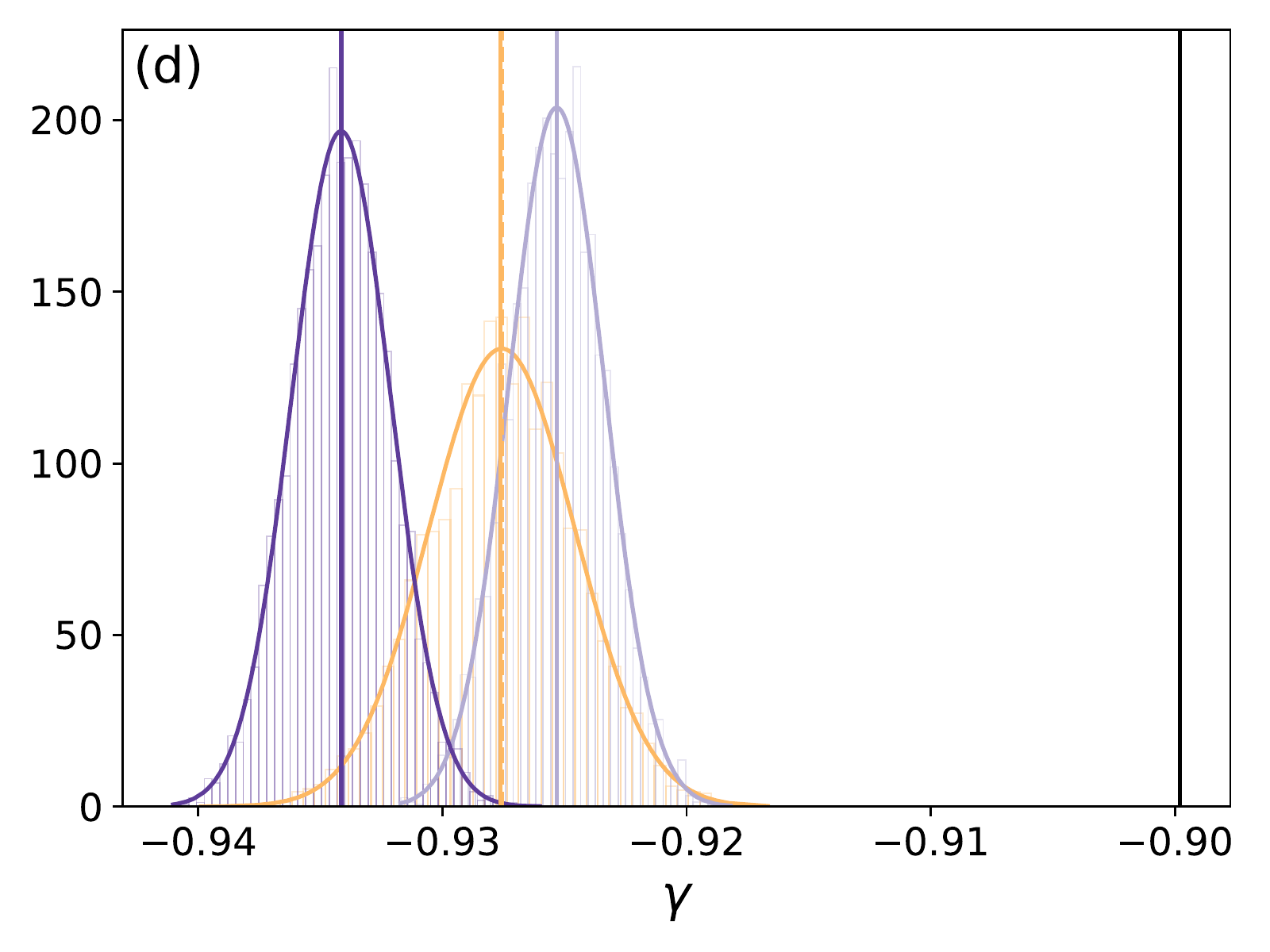}
\caption{Estimations of the parameters (a) $\beta_0$, (b) $\rho$, (c) $\alpha$ and (d) $\gamma$, entering the definition of the econometric version of the CEM, where the binary topology is either `deterministic' (black vertical line) or generated via the UBRGM (light orange or light grey), the UBCM (purple or dark grey) and the LM (light purple or grey). The deterministic approach leads to a single estimate, while the other approaches lead to either a single, `annealed' estimate (vertical, solid lines) or to a whole distribution of `quenched' estimates (histograms with kernel density curves, constructed over an ensemble of 5.000 binary configurations; the corresponding average value is indicated by a vertical, dash-dotted line). Each `annealed' parameter estimate coincides with the average value of the corresponding `quenched' distribution although the distributions induced by the three, binary recipes may overlap or not; the `deterministic' estimate, instead, overlaps with the other, two ones only for the parameter $\alpha$, under the UBCM-induced, binary recipe. Data refers to the year 2017 of the CEPII-BACI version of the WTW~\cite{Baci2014}.}
\label{fig3}
\end{figure*}

The `annealed' recipe, instead, prescribes to maximise the likelihood

\begin{align}
\mathcal{G}_{\underline{\psi}}&=\sum_{i<j}[-(\beta_i+\beta_j)w_{ij}^*+p_{ij}\ln(\beta_i+\beta_j)]\nonumber\\
&=-\sum_i\beta_is_i^*+\sum_{i<j}p_{ij}\ln(\beta_i+\beta_j)
\end{align}
whose optimisation requires to solve the system of equations

\begin{equation}\label{eqs}
s_i^*=\sum_{j(\neq i)}\frac{p_{ij}}{\beta_i+\beta_j},\:\forall\:i
\end{equation}
(notice that both the `deterministic' and the `annealed' version of the `vector' variant of the CEM are alternative instances of the so-called $\text{CReM}_\text{A}$, introduced in~\cite{Parisi2020}). The `quenched' recipe, on the other hand, requires to solve the system of equations $\langle\beta_i\rangle=\sum_{\mathbf{A}\in\mathbb{A}}P(\mathbf{A})\beta_i(\mathbf{A})$, $\forall\:i$ which no longer have an explicit expression. Devising some sort of approximation is, however, possible. Let us start by re-writing eq. \ref{eqs} as

\begin{equation}
\beta_i=\frac{1}{s_i^*}\sum_{j(\neq i)}\frac{p_{ij}}{1+\beta_j/\beta_i},\:\forall\:i
\end{equation}
and consider the node whose coefficient is the largest one. This allows us to write $\beta_i\simeq\sum_{j(\neq i)}p_{ij}/s_i^*=\langle k_i\rangle/s_i^*$: in case we implemented the UBRGM, we would obtain $\beta_i(\mathbf{A})\simeq 2L(\mathbf{A})/Ns_i^*$, hence expecting the `quenched' distribution of $Ns_i^*\beta_i/2$ to coincide with $\text{Bin}(N(N-1)/2,p)$; if, on the other hand, we implemented the UBCM, we would obtain $\beta_i(\mathbf{A})\propto k_i(\mathbf{A})/s_i^*$, hence expecting the `quenched' distribution of $s_i^*\beta_i$ to obey a Poisson-Binomial. Again, the estimations coincide for any null model preserving the structural properties characterising the binary recipe implemented.

More generally, the mutual relationships between the estimations provided by the three recipes are node-dependent (see Fig.~\ref{fig2}, illustrating the case-study of node 166 of the WTW and Fig.~\ref{figD1} in Appendix C): in general, however, each `annealed' estimation overlaps with the average value of the related `quenched' distribution. Moreover, the `deterministic' estimation is very close to the UBCM-induced, `annealed' one; such a result is a consequence of the accurate description of the empirical network topology provided by the UBCM - in fact, much more accurate than the ones provided by the UBRGM and the LM: indeed, the better the approximation $p_{ij}\simeq a_{ij}$, $\forall\:i<j$, the closer the `annealed' estimation to the `deterministic' one.

This is even more evident when considering the `tensor' variant of the CEM, in which case the three optimisation procedures lead to the expressions $\beta_\text{det}=a_{ij}^*/\hat{w}_{ij}$, $\forall\:i<j$ and $\beta_\text{ann}=\langle\beta\rangle_\text{que}=p_{ij}/\hat{w}_{ij}$, $\forall\:i<j$ - with $\hat{w}_{ij}$ representing an estimate of the empirical weight $w_{ij}^*$; if, however, $\hat{w}_{ij}\equiv w_{ij}^*$, $\forall\:i<j$ then, for consistency, $p_{ij}\equiv a_{ij}^*$ and the three recipes coincide.

\subsection{`Econometric' variant of the\\Conditional Exponential Model}

As a third case-study, let us focus on the `econometric' variant of the CEM, defined by posing $\beta_{ij}\equiv\beta_0+z_{ij}^{-1}$, $\forall\:i<j$, where $z_{ij}\equiv e^{\rho}(\omega_i\omega_j)^\alpha d_{ij}^\gamma$ represents the Gravity Model specification traditionally employed to analyse undirected, weighted, trade networks and $\beta_0$ is a structural parameter to be tuned in order to ensure that $\langle W\rangle=W^*$. In this case, the `deterministic' recipe for parameter estimation prescribes to maximise the likelihood

\begin{align}
\mathcal{L}_{\underline{\psi}}&=\sum_{i<j}[-(\beta_0+z_{ij}^{-1})w_{ij}^*+a_{ij}^*\ln(\beta_0+z_{ij}^{-1})]
\end{align}
whose optimisation requires to solve the system of equations

\begin{align}
W^*&=\sum_{i<j}\frac{a_{ij}^*}{\beta_0+z_{ij}^{-1}},\\
\sum_{i<j}w_{ij}^*\cdot\frac{\partial z_{ij}^{-1}}{\partial\underline{\phi}}&=\sum_{i<j}\frac{a_{ij}^*}{\beta_0+z_{ij}^{-1}}\cdot\frac{\partial z_{ij}^{-1}}{\partial\underline{\phi}}.
\end{align}

The `annealed' recipe, instead, prescribes to maximise the likelihood

\begin{align}
\mathcal{G}_{\underline{\psi}}=\sum_{i<j}[-(\beta_0+z_{ij}^{-1})w_{ij}^*+p_{ij}\ln(\beta_0+z_{ij}^{-1})]
\end{align}
whose optimisation requires to solve the system of equations

\begin{align}
W^*&=\sum_{i<j}\frac{p_{ij}}{\beta_0+z_{ij}^{-1}},\\
\sum_{i<j}w_{ij}^*\cdot\frac{\partial z_{ij}^{-1}}{\partial\underline{\phi}}&=\sum_{i<j}\frac{p_{ij}}{\beta_0+z_{ij}^{-1}}\cdot\frac{\partial z_{ij}^{-1}}{\partial\underline{\phi}}.
\end{align}

The `quenched' recipe, on the other hand, requires to solve the system of equations $\langle\beta_{0}\rangle=\sum_{\mathbf{A}\in\mathbb{A}}P(\mathbf{A})\beta_{0}(\mathbf{A})$ and $\langle\underline{\phi}\rangle=\sum_{\mathbf{A}\in\mathbb{A}}P(\mathbf{A})\underline{\phi}(\mathbf{A})$ which no longer have an explicit expression.

Figures~\ref{fig3} and \ref{figD2} in Appendix C illustrate the case-study of the WTW: although the `quenched' distributions induced by the three, binary recipes are characterised by different shapes that may overlap (as in the case of the parameters $\rho$ - under the UBRGM-induced and UBCM-induced binary recipes - and $\gamma$ - under all, binary recipes) or not (as in the case of the parameters $\beta_0$ and $\alpha$), `annealed' and `quenched' estimations always coincide (the only, small discrepancy being observable for the parameter $\beta_0$, under the UBRGM-induced, binary recipe). The `deterministic' estimation, instead, is compatible with the other, two ones only for the parameter $\alpha$, under the UBCM-induced, binary recipe.\\

Sparse networks deserve a separate discussion. The results concerning the homogeneous and econometric variant of the BLN, defined by posing $\beta_{ij}\equiv\beta_0+z_{ij}^{-1}$, $\forall\:i<j$, with $z_{ij}\equiv e^{\rho}(s_i s_j)^\alpha$, are analogous to the ones shown for the WTW - in the latter case, the `annealed' estimates of $\beta_0$, $\rho$ and $\alpha$ are very close to their `quenched' counterparts, the relative error $\text{RE}=|(\phi_i^\text{ann}-\phi_i^\text{que})/\phi_i^\text{ann}|$ amounting at $\simeq 10^{-3}$ for $\beta_0$ and $\simeq 10^{-4}$ for $\rho$, $\alpha$. On the contrary, these conclusions no longer hold true when the weakly heterogeneous variant of the CEM is considered: in this case, in fact, carrying out the `quenched' approach can lead to binary configurations with disconnected nodes, a circumstance that impairs the correct estimation of the corresponding parameters; carrying out the `annealed' estimation, instead, remains a feasible task.

\section{Discussion}

The present contribution focuses on three recipes for estimating the parameters entering into the definition of statistical network models, i.e. the `deterministic', `annealed' and `quenched' ones. In order to implement them, we have considered several variants of the CEM, i.e. the homogeneous one (defined by one, global parameter), the weakly heterogeneous one (defined by $N$, local parameters) and the econometric one (defined by four, global parameters), each one combined with three, different recipes for estimating the network topology (i.e. the UBRGM, the UBCM and the LM).

The `deterministic' recipe, routinely employed in econometrics to determine the so-called hurdle models~\cite{Mullahy}, prescribes to estimate the parameters associated to the weighted constraints on the empirical realisation of the network topology. Since it considers $\mathbf{A}^*$ as not being subject to variation, its use is recommended whenever $\text{Var}[a_{ij}]=p_{ij}(1-p_{ij})\simeq 0$ or, equivalently, $p_{ij}\simeq a_{ij}$, $\forall\:i<j$, i.e. whenever the binary random variables can be safely considered as deterministic or, more in general, whenever their (scale of) variation is negligible with respect to the (scale of) variation of the weighted random variables.

Accounting for such a variability in a fully consistent manner can be achieved upon adopting either the `annealed' recipe (according to which parameters are estimated on the average network topology) or the `quenched' recipe (according to which parameters are, first, estimated on a large number of binary configurations and, then, averaged); the main difference between these procedures lies in the order in which the two operations of `averaging' (of the entries of the binary adjacency matrix) and `maximisation' (of the related likelihood function) are taken. Interestingly, no variant of the CEM is sensitive to this choice (neither the purely structural ones nor the `econometric' one); while, however, the coincidence of the `annealed' and `quenched' estimates for purely structural models can be explicitly verified, this is no longer true when the `econometric' variant is considered: in this case, in fact, one can proceed only numerically.

This evidence reveals the main limitation of the `quenched' approach, i.e. the need of resorting upon an explicit sampling of the chosen, binary ensemble. As any `good' sampling algorithm must lead to a faithful representation of the parent distribution, we are left with the following question: \emph{is this always guaranteed, in all cases of interest to us?}

This seems to be the case for dense networks. As shown in \cite{Tiziano2015}, a study of the coefficient of variation of the constraints defining the `vector' variant of the CEM (i.e. the ratio between standard deviation and expected value of each degree) reveals it to vanish in the asymptotic limit: in other words, the fluctuations affecting each degree vanish, a result guaranteeing that the degree sequence of any configuration in the ensemble remains `close enough' to the empirical one.

When sparse networks are, instead, considered, the coefficient of variation of the constraints defining the `vector' variant of the CEM remains finite in the asymptotic limit: in other words, the fluctuations affecting each degree do not vanish, a result implying that the degree sequence of any configuration in the ensemble may largely differ from the empirical one; to provide a concrete example, nodes whose empirical degree is `small' may disconnect, hence inducing the resolution of a system of equations which is not even compatible with the set of constraints defining the original problem. Overcoming such a limitation implies quantifying the bias affecting the estimates in cases like these: although possible, calculations of this kind are far beyond the scope of the present paper.\\

Overall, then, two alternatives exist to overcome the main limitation of the `deterministic' estimation recipe, i.e. that of ignoring the variety of structures that are compatible with a given probability distribution $P(\mathbf{A})$, namely the `annealed' and `quenched' ones. As the `quenched' recipe requires an explicit sampling the ensemble - potentially leading to inconsistent estimates for sparse configurations - we believe the `annealed' one to represent the better alternative, 1) being \emph{unbiased} by definition, 2) being \emph{convenient} from a numerical point of view, 3) reducing to the `deterministic' recipe in case the empirical configuration is not subject to variation.

\section{Acknowledgements}

SoBigData.it receives funding from European Union – NextGenerationEU – National Recovery and Resilience Plan (Piano Nazionale di Ripresa e Resilienza, PNRR) – Project: “SoBigData.it – Strengthening the Italian RI for Social Mining and Big Data Analytics” – Prot. IR0000013 – Avviso n. 3264 del 28/12/2021. This work is also supported by PNRR-M4C2-Investimento 1.3, Partenariato Esteso PE00000013 - `FAIR-Future Artificial Intelligence Research' - Spoke 1 `Human-centered AI', funded by the European Commission under the NextGeneration EU programme and by the project `Network analysis of economic and financial resilience', Italian DM n. 289, 25-03-2021 (PRO3 Scuole) CUP D67G22000130001.
DG acknowledges support from the Dutch Econophysics Foundation (Stichting Econophysics, Leiden, the Netherlands) and the Netherlands Organization for Scientific Research (NWO/OCW).
MDV acknowledges support from the European Union ERC-2018-ADG Grant Agreement n. 834756, `XAI: Science and technology for the explanation of AI decision making'.
MDV and DG also acknowledge support from the `Programma di Attività Integrata' (PAI) project `Prosociality, Cognition and Peer Effects' (Pro.Co.P.E.), funded by IMT School for Advanced Studies Lucca.

\section*{Appendix A.\\Conditional network models\\from KL divergence minimisation}

Discrete maximum-entropy models can be derived by performing a constrained maximisation of Shannon entropy~\cite{Shannon,Jaynes1957a}. Here, however, we focus on continuous probability distributions: in such a case, mathematical problems are known to affect the definition of Shannon entropy as well as the resulting inference procedure: to restore the framework, one has to consider the KL divergence $D_\text{KL}(Q||R)$ of a distribution $Q(\mathbf{W})$ from a prior distribution $R(\mathbf{W})$ and re-interpret the maximisation of the entropy associated to $Q(\mathbf{W})$ as the minimisation of its `distance' from $R(\mathbf{W})$. Such an optimisation scheme embodies the so-called Minimum Discrimination Information Principle, originally proposed by Kullback and Leibler~\cite{Kullback1951} and requiring new data to produce an information gain that is as small as possible. In formulas, the KL divergence is defined as

\begin{equation}
D_\text{KL}(Q||R)=\int_\mathbb{W}Q(\mathbf{W})\ln \frac{Q(\mathbf{W})}{R(\mathbf{W})}d\mathbf{W};
\end{equation}
the class of conditional models can be introduced upon re-writing the posterior distribution $Q(\mathbf{W})$ as $Q(\mathbf{W})=P(\mathbf{A})Q(\mathbf{W}|\mathbf{A})$, where $\mathbf{A}$ denotes the binary projection of the weighted network $\mathbf{W}$. This equation allows us to split the KL divergence into the sum of three terms reading

\begin{equation}
D_\text{KL}(Q||R)=S(Q,R)-S(P)-S(\overline{Q}|P)
\end{equation}
where

\begin{equation}
S(Q,R)=-\sum_{\mathbf{A}\in\mathbb{A}}P(\mathbf{A})\int_{\mathbb{W}_\mathbf{A}}Q(\mathbf{W}|\mathbf{A})\ln R(\mathbf{W})d\mathbf{W}
\end{equation}
is the cross entropy, quantifying the amount of information required to identify a weighted network sampled from the distribution $Q(\mathbf{W})$ by employing the distribution $R(\mathbf{W})$,

\begin{equation}
S(P)=-\sum_{\mathbf{A}\in\mathbb{A}}P(\mathbf{A})\ln P(\mathbf{A})
\end{equation}
is the Shannon entropy of the probability distribution describing the binary projection of the network structure and

\begin{equation}
S(\overline{Q}|P)=-\sum_{\mathbf{A}\in\mathbb{A}}P(\mathbf{A})\int_{\mathbb{W}_\mathbf{A}}Q(\mathbf{W}|\mathbf{A})\ln Q(\mathbf{W}|\mathbf{A})d\mathbf{W}
\end{equation}
is the conditional Shannon entropy of the probability distribution of the weighted network structure, given the binary projection. The expression for $S(Q,R)$ can be further manipulated as follows: upon separating the prior distribution itself into a purely binary part and a conditional, weighted one, we can pose $R(\mathbf{W})=T(\mathbf{A})R(\mathbf{W}|\mathbf{A})$, an expression that allows the KL divergence to be re-written as 

\begin{equation}
D_\text{KL}(Q||R)= D_\text{KL}(P||T) + D_\text{KL}(\overline{Q}||\overline{R})
\end{equation}
i.e. as a sum of the two addenda

\begin{align}
D_\text{KL}(P||T)&=\sum_{\mathbf{A}\in\mathbb{A}}P(\mathbf{A})\ln\frac{P(\mathbf{A})}{T(\mathbf{A})},\\
D_\text{KL}(\overline{Q}||\overline{R})&=\sum_{\mathbf{A}\in\mathbb{A}}P(\mathbf{A})\int_{\mathbb{W}_\mathbf{A}}Q(\mathbf{W}|\mathbf{A})\ln\frac{Q(\mathbf{W}|\mathbf{A})}{R(\mathbf{W}|\mathbf{A})}d\mathbf{W}
\end{align}
with $T(\mathbf{A})$ representing the binary prior and $R(\mathbf{W}|\mathbf{A})$ representing the conditional, weighted one. Dealing with completely uninformative priors amounts at considering the expression

\begin{equation}
-S(Q)=-S(P)-S(\overline{Q}|P)
\end{equation}
i.e. `minus' the joint entropy. The (independent) constrained optimisation of $S(P)$ and $S(\overline{Q}|P)$ represents the starting point for deriving the members of the class of conditional models.

\section*{Appendix B.\\Conditional network models: determining the functional form}

The constrained maximisation of $S(\overline{Q}|P)$ proceeds by specifying the set of weighted constraints reading

\begin{align}
1&=\int_{\mathbb{W}_\mathbf{A}}P(\mathbf{W}|\mathbf{A})d\mathbf{W},\:\forall\:\mathbf{A}\in\mathbb{A},\\
\langle C_\alpha\rangle&=\sum_{\mathbf{A}\in\mathbb{A}}P(\mathbf{A})\int_{\mathbb{W}_\mathbf{A}}Q(\mathbf{W}|\mathbf{A})C_\alpha(\mathbf{W})d\mathbf{W},\:\forall\:\alpha
\end{align}
the first condition ensuring the normalisation of the probability distribution and the vector $\{C_\alpha(\mathbf{W})\}$ representing the proper set of weighted constraints. The distribution induced by them reads

\begin{align}
Q(\mathbf{W}|\mathbf{A}) &= \frac{e^{-H(\mathbf{W})}}{Z_\mathbf{A}}=\frac{e^{-H(\mathbf{W})}}{\int_{\mathbb{W}_\mathbf{A}}e^{-H(\mathbf{W})}d\mathbf{W}}= \nonumber \\
&=\frac{e^{-\sum_{i<j}H_{ij}(w_{ij})}}{\int_{\mathbb{W}_\mathbf{A}}e^{-\sum_{i<j}H_{ij}(w_{ij})}d\mathbf{W}}= \nonumber \\
&=\prod_{i<j}\frac{e^{-H_{ij}(w_{ij})}}{\left[\int_{m_{ij}}^{+\infty}e^{-H_{ij}(w_{ij})}dw_{ij}\right]^{a_{ij}}}=\prod_{i<j}\frac{e^{-H_{ij}(w_{ij})}}{\zeta_{ij}^{a_{ij}}}
\end{align}
if $\mathbf{W}\in\mathbb{W}_\mathbf{A}$ and $0$ otherwise - since each Hamiltonian considered in the present paper is separable, i.e. a sum of node pairs-specific Hamiltonians: in formulas, $H(\mathbf{W})=\sum_{i<j}H_{ij}(w_{ij})$.

\section*{Appendix C.\\Conditional network models: estimating the parameters}

Let us, now, provide general expressions for the `deterministic' and the `annealed' recipe for parameter estimation. The first one follows from writing

\begin{align}
\mathcal{L}_{\underline{\psi}}&=\ln Q(\mathbf{W}^*|\mathbf{A}^*)=-H(\mathbf{W}^*)-\ln Z_{\mathbf{A}^*}= \nonumber \\
&= - H(\mathbf{W}^*)-\ln\left[\int_{\mathbb{W}_{\mathbf{A}^*}}e^{-H(\mathbf{W})}d\mathbf{W}\right] = \nonumber\\
&=\sum_{i<j}H_{ij}(w_{ij}^*)-\ln\prod_{i<j}\zeta_{ij}^{a_{ij}}=\sum_{i<j}[H_{ij}(w_{ij}^*)-a_{ij}^*\ln\zeta_{ij}]
\end{align}
while the second one follows from writing

\begin{align}
\mathcal{G}_{\underline{\psi}}&=\sum_{\mathbf{A}\in\mathbb{A}}P(\mathbf{A})\ln Q(\mathbf{W}^*|\mathbf{A})=\langle\mathcal{L}_{\underline{\psi}}\rangle= \nonumber \\
&= \sum_{i<j}[H_{ij}(w_{ij}^*)-p_{ij}\ln\zeta_{ij}].
\end{align}

\paragraph*{`Scalar' or homogeneous variant of the CEM.} In the particular case of the UBRGM-induced, homogeneous variant of the CEM, one can derive the `quenched' distribution of the parameter $\beta$ upon considering that it is a function of the discrete, random variable $L$. Since $L\sim\text{Bin}(N(N-1)/2,p)$, with $p=2L^*/N(N-1)$, one finds that

\begin{equation}
\beta\sim\binom{\frac{N(N-1)}{2}}{W^*\beta}p^{W^*\beta}(1-p)^{\frac{N(N-1)}{2}-W^*\beta}
\end{equation}
an expression allowing us to derive the expected value of $\beta$, i.e.

\begin{align}
\langle\beta\rangle &= \sum_{\beta=0}^{\frac{N(N-1)}{2W^*}}\beta\binom{\frac{N(N-1)}{2}}{W^*\beta}p^{W^*\beta}(1-p)^{\frac{N(N-1)}{2}-W^*\beta}= \nonumber \\
&= \frac{N(N-1)}{2W^*}p=\frac{\langle L\rangle}{W^*}=\frac{L^*}{W^*}
\end{align}
as well as its variance. Since 

\begin{align}
\langle\beta^2\rangle &=\sum_{\beta=0}^{\frac{N(N-1)}{2W^*}}\beta^2\binom{\frac{N(N-1)}{2}}{W^*\beta}p^{W^*\beta}(1-p)^{\frac{N(N-1)}{2}-W^*\beta}= \nonumber \\
&= \frac{N(N-1)}{2(W^*)^2}p+\frac{N(N-1)}{2(W^*)^2}\left[\frac{N(N-1)}{2(W^*)^2}-1\right]p^2
\end{align}
we have that

\begin{align}
\text{Var}[\beta] &=\langle\beta^2\rangle-\langle\beta\rangle^2=\frac{N(N-1)}{2(W^*)^2}p(1-p)=\frac{\text{Var}[L]}{(W^*)^2}= \nonumber \\
&= \frac{L^*}{(W^*)^2}\left[\frac{N(N-1)-2L^*}{N(N-1)}\right]
\end{align}
with $\text{Var}[L]=N(N-1)/2\cdot p(1-p)$. Since the distribution obeyed by $L$ converges to the normal distribution $\mathcal{N}(L^*,\text{Var}[L])$, the distribution obeyed by $\beta$ converges to the distribution

\begin{align}
g(\beta)&=\frac{W^*}{\sqrt{2\pi\text{Var}[L]}}e^{-\frac{(W^*\beta-L^*)^2}{2\text{Var}[L]}}=\nonumber \\ &= \frac{1}{\sqrt{2\pi\text{Var}[L]/(W^*)^2}}e^{-\frac{(\beta-L^*/W^*)^2}{2\text{Var}[L]/(W^*)^2}}= \nonumber \\
&= \frac{1}{\sqrt{2\pi\text{Var}[\beta]}}e^{-\frac{(\beta-\beta^*)^2}{2\text{Var}[\beta]}}=\mathcal{N}(\beta^*,\text{Var}[\beta])
\end{align}
with $\beta^*=L^*/W^*$ and $\text{Var}[\beta]=\text{Var}[L]/(W^*)^2$.\\

\begin{figure*}[t!]
\centering
\includegraphics[width=0.32\textwidth]{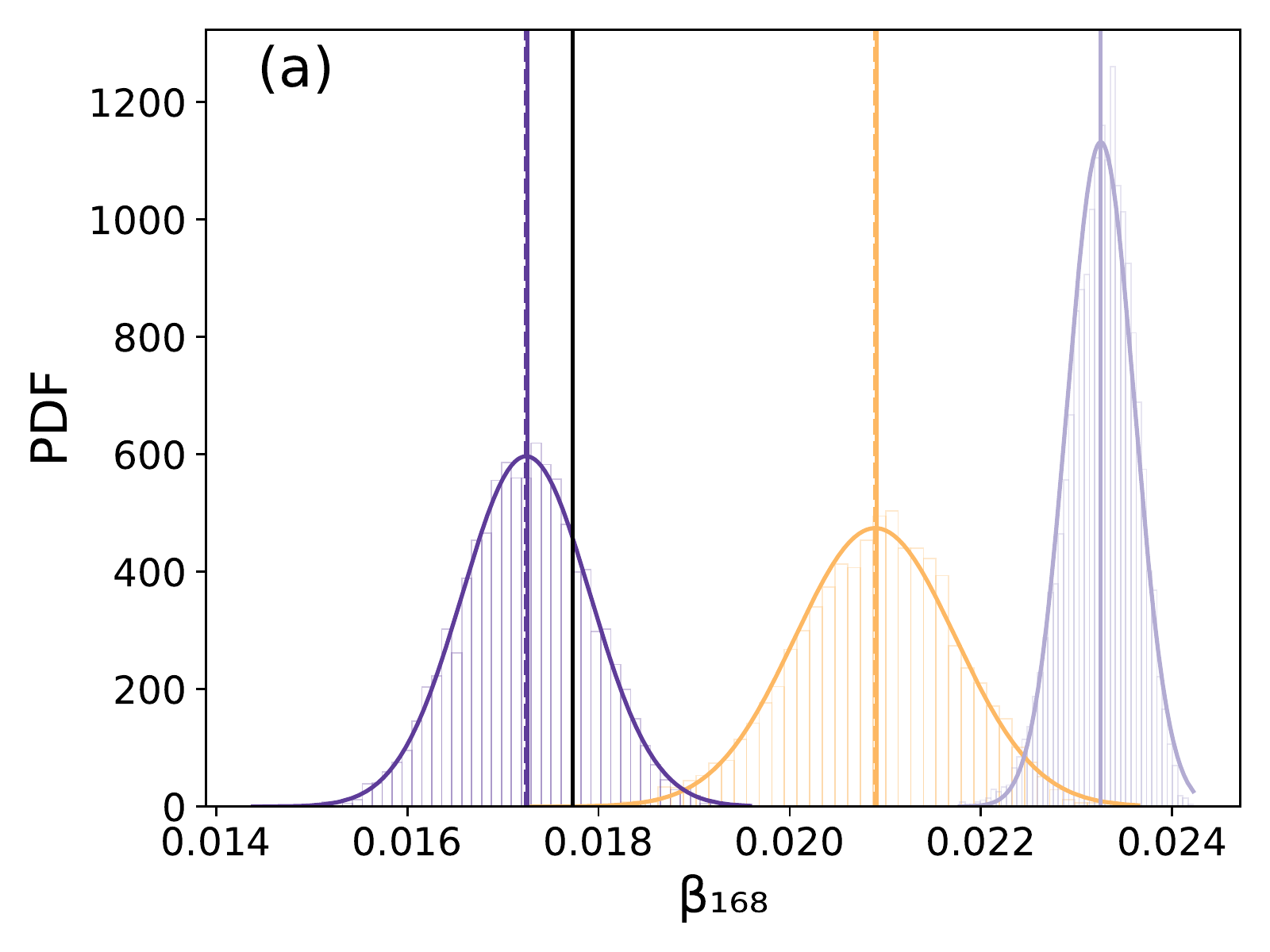}
\includegraphics[width=0.32\textwidth]{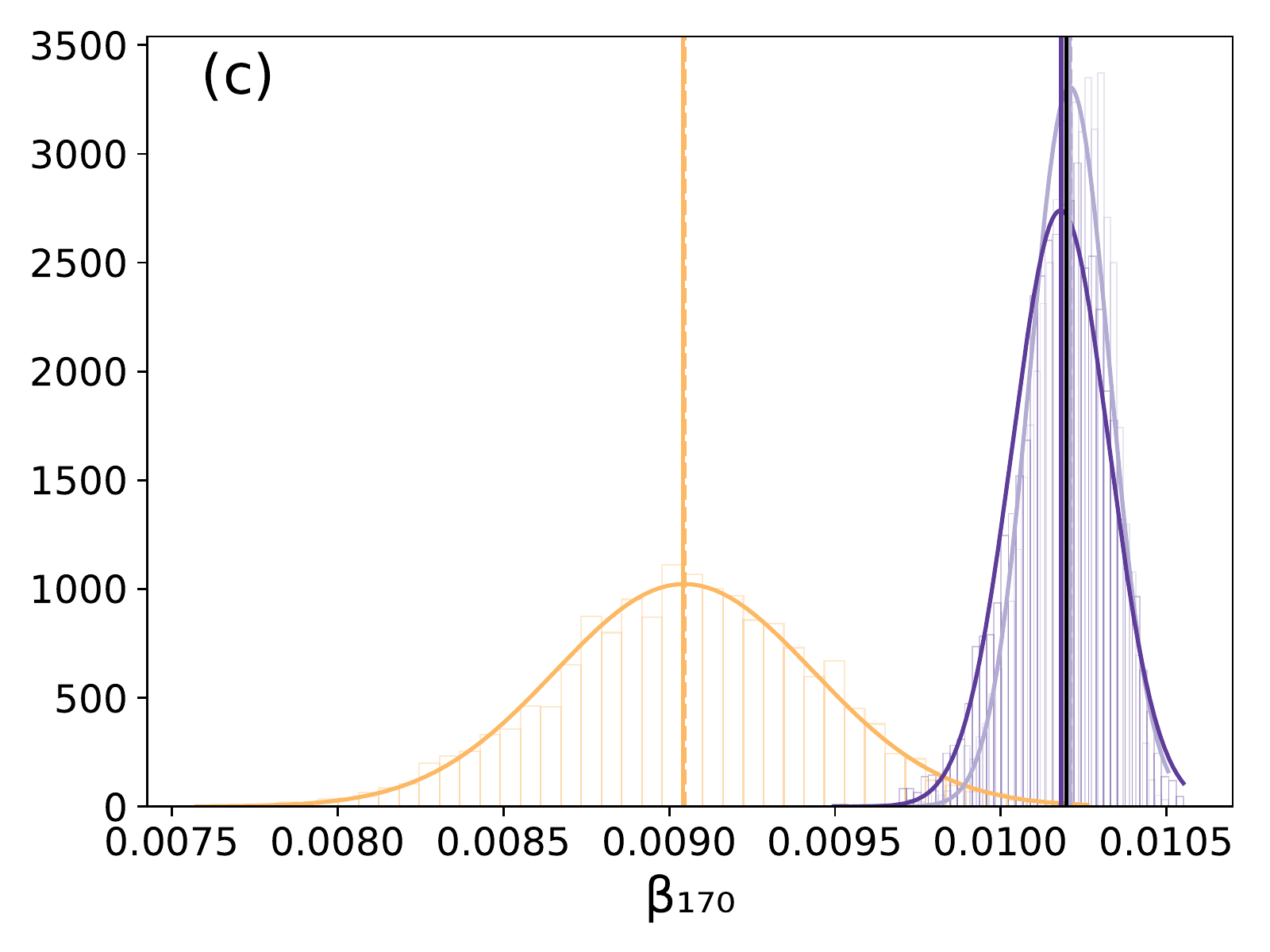}
\includegraphics[width=0.32\textwidth]{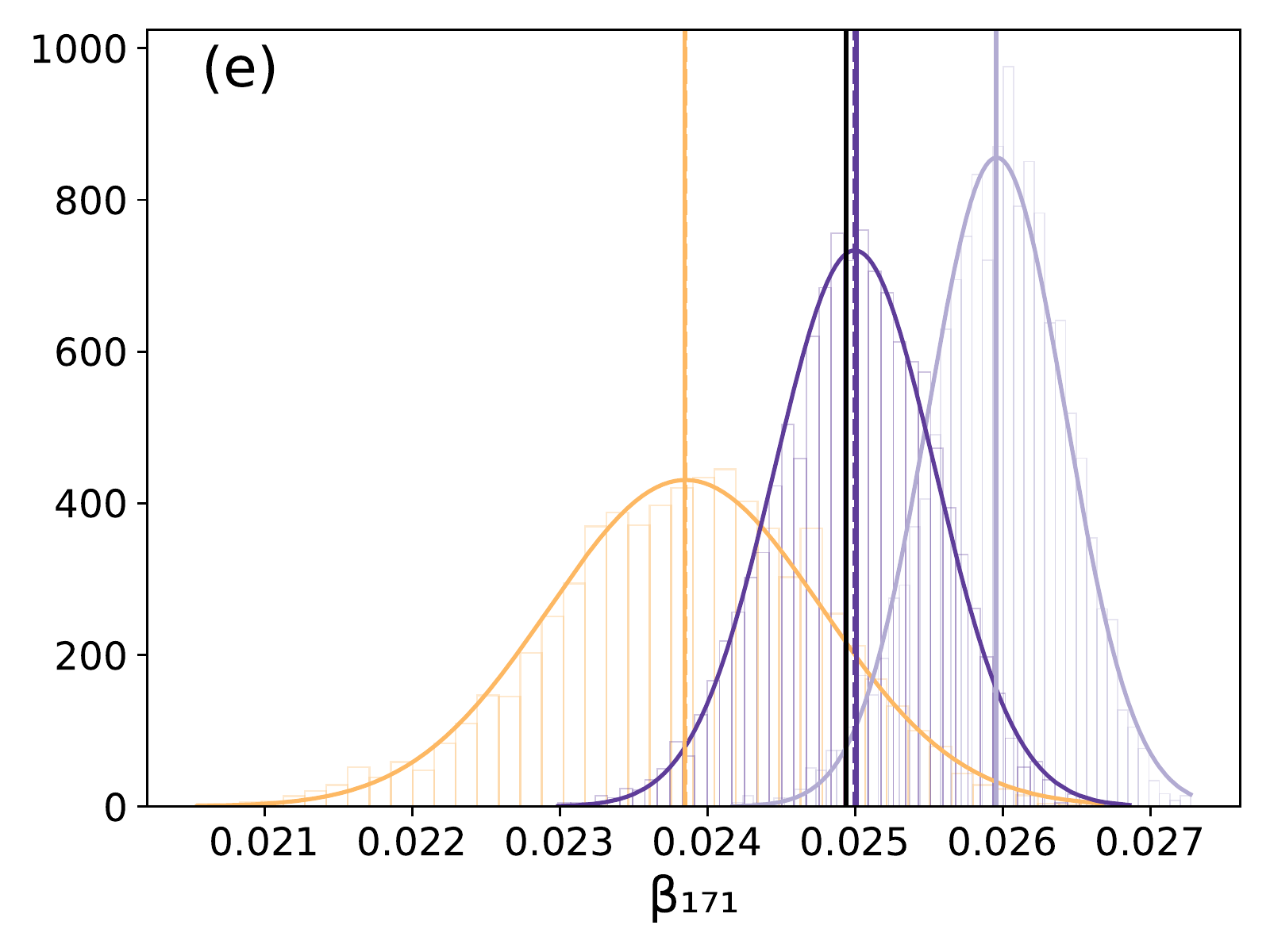}\\
\includegraphics[width=0.32\textwidth]{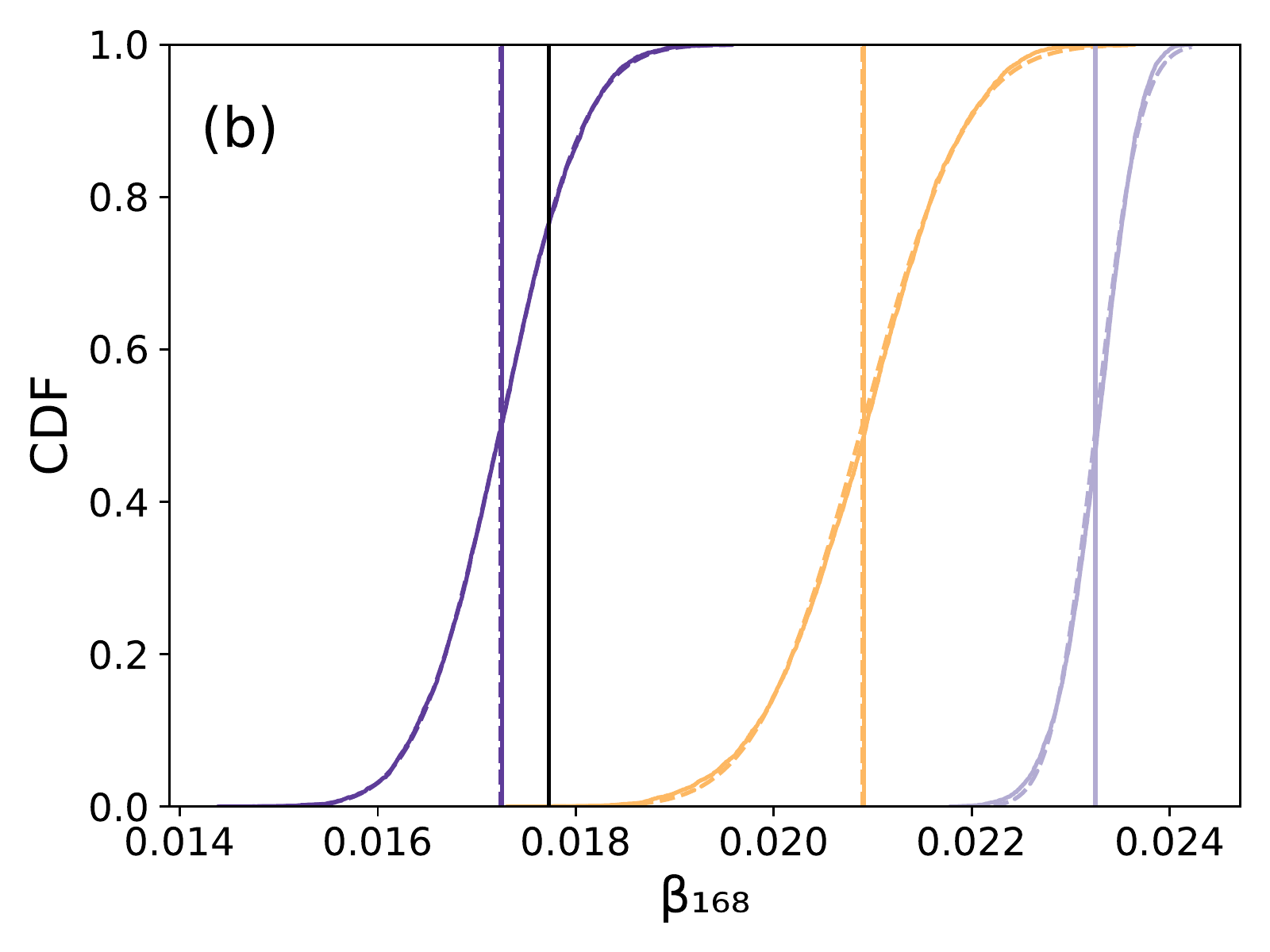}
\includegraphics[width=0.32\textwidth]{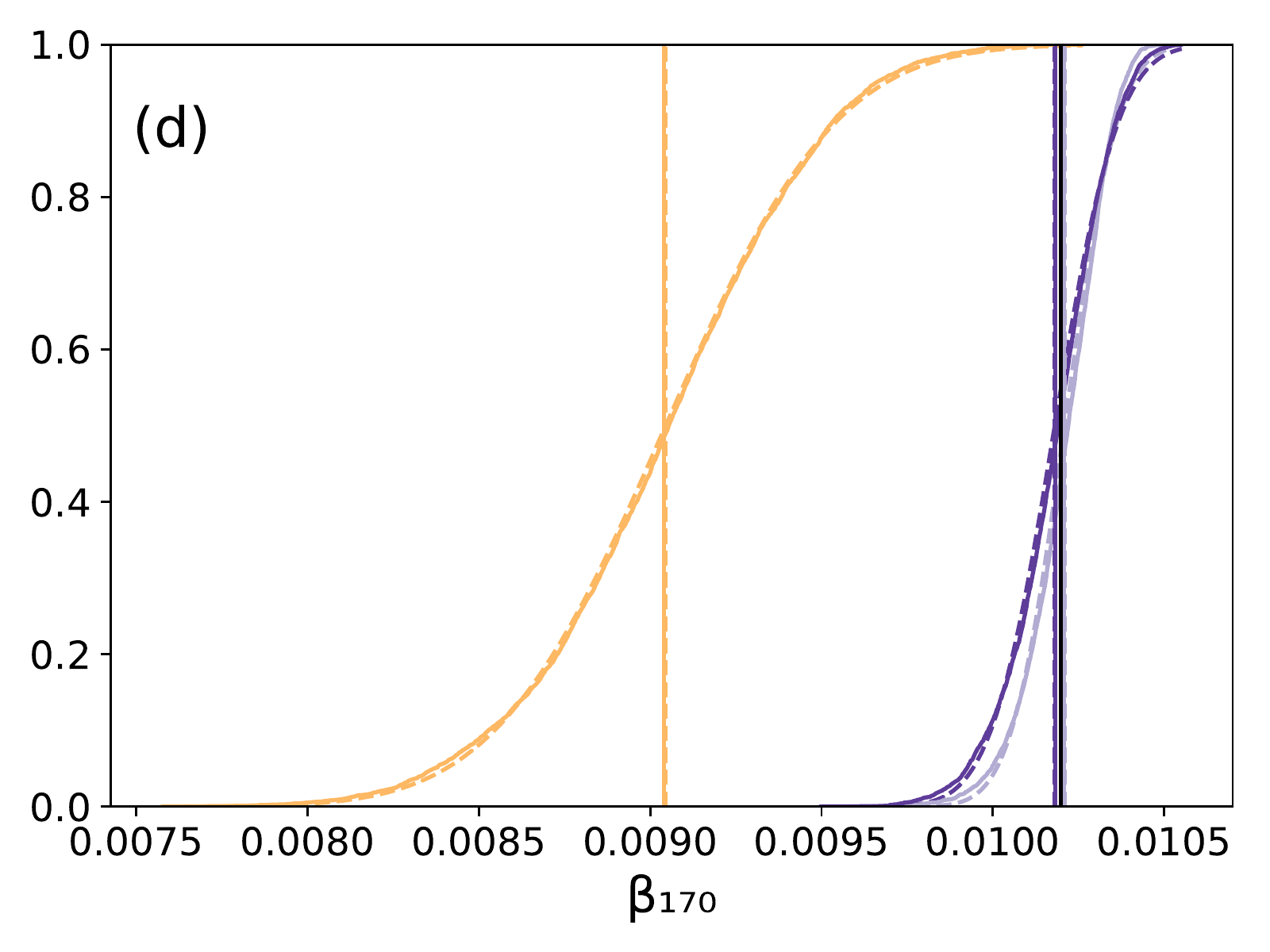}
\includegraphics[width=0.32\textwidth]{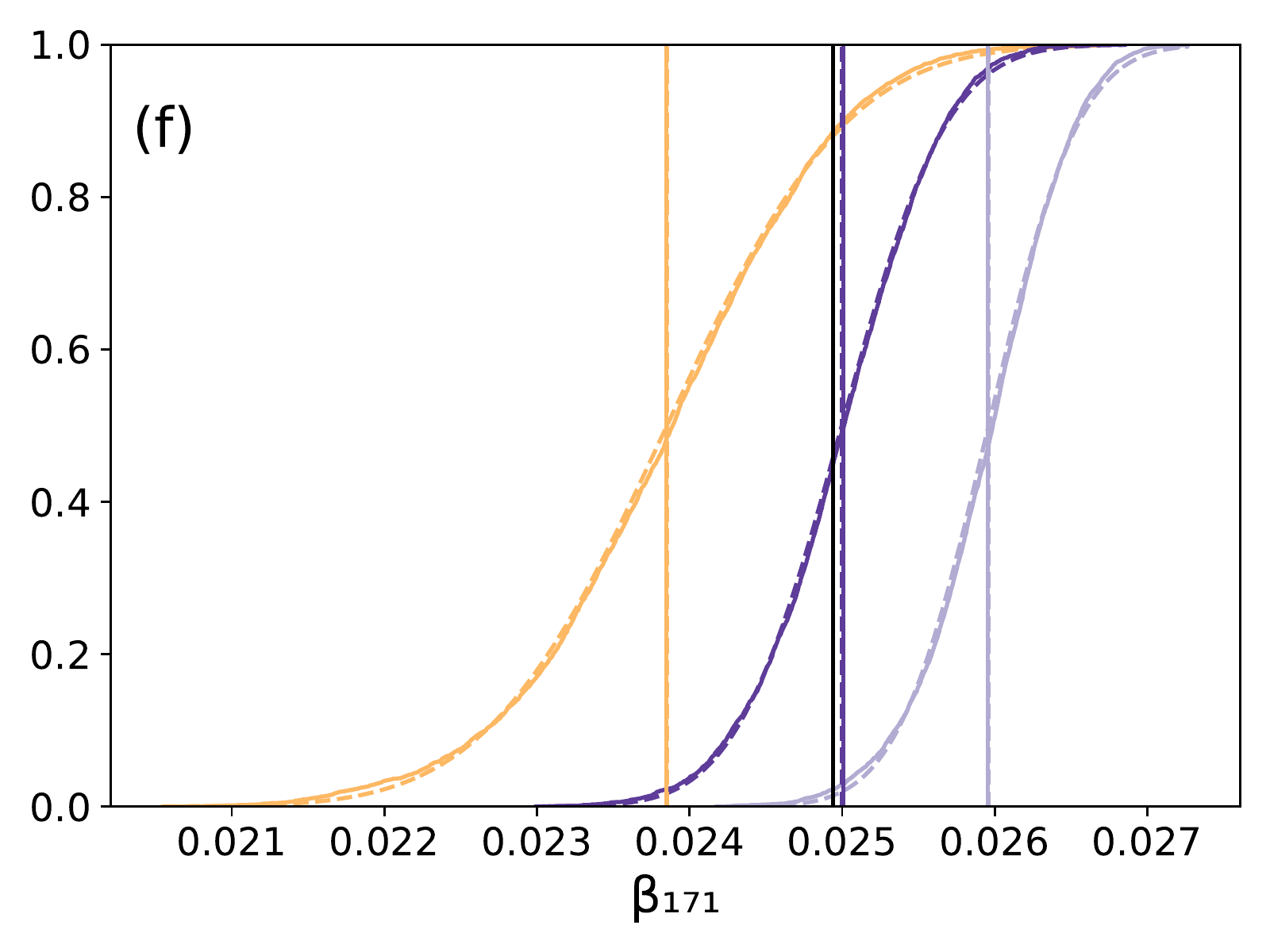}
\caption{Estimations of the parameters (a)-(b) $\beta_{168}$ , (c)-(d) $\beta_{170}$ and (e)-(f) $\beta_{171}$, entering the definition of the weakly heterogeneous version of the CEM, where the binary topology is either `deterministic' (black vertical line) or generated via the UBRGM (light orange or light grey), the UBCM (purple or dark grey) and the LM (light purple or grey). The deterministic approach leads to a single estimate, while the other approaches lead to either a single, `annealed' estimate (vertical, solid lines) or to a whole distribution of `quenched' estimates (histograms with normal density curves having the same average and standard deviation, constructed over an ensemble of 5.000 binary configurations; the average value is indicated by a vertical, dash-dotted line). Each `annealed' estimate overlaps with the average value of the related `quenched' distribution, although 1) the latter ones are well separated in the case of node 168, 2) only partly overlapped in the case of node 171, 3) the UBCM-induced and the LM-induced ones overlap while the UBRGM-induced one remains well separated in the case of node 170. Moreover, the `deterministic' estimates are always very close to (if not overlapping with) the UBCM-induced, `annealed' ones. Although the empirical and theoretical CDFs (respectively depicted as solid lines and dotted lines in the bottom panels) seem to be in a very good agreement, the Anderson-Darling test never rejects the normality hypothesis only for node 166 and does not reject the normality hypothesis in the case of the UBCM-induced distribution of estimates for node 168.}
\label{figD1}
\end{figure*}

In the case of the UBCM-induced, homogeneous version of the CEM, $L$ obeys the Poisson-Binomial (PB) distribution reading $\text{PB}(N(N-1)/2,\{p^\text{UBCM}\}_{i,j=1}^N)$ whose normal approximation reads $\mathcal{N}(L^*,\text{Var}[L])$, with $\text{Var}[L]=\sum_{i<j}p^\text{UBCM}_{ij}(1-p^\text{UBCM}_{ij})$; as a consequence, the distribution obeyed by $\beta$ converges to $\mathcal{N}(\beta^*,\text{Var}[\beta])$, with $\beta^*=L^*/W^*$ and $\text{Var}[\beta]=\text{Var}[L]/(W^*)^2$.

In the case of the LM-induced, homogeneous version of the CEM, $L$ obeys the Poisson-Binomial distribution reading $\text{PB}(N(N-1)/2,\{p^\text{LM}\}_{i,j=1}^N)$ whose normal approximation reads $\mathcal{N}(L^*,\text{Var}[L])$, with $\text{Var}[L]=\sum_{i<j}p^\text{LM}_{ij}(1-p^\text{LM}_{ij})$; as a consequence, the distribution obeyed by $\beta$ converges to $\mathcal{N}(\beta^*,\text{Var}[\beta])$, with $\beta^*=L^*/W^*$ and $\text{Var}[\beta]=\text{Var}[L]/(W^*)^2$.\\

\paragraph*{`Vector' or weakly heterogeneous variant of the CEM.} As pointed out in the main text, each `annealed' estimation overlaps with the average value of the related `quenched' distribution although 1) the latter ones are well separated, in the case of node 168, 2) only partly overlapped, in the case of node 171, 3) the UBCM-induced and the LM-induced ones overlap while the UBRGM-induced one remains well separated, in the case of node 170 (see Fig.~\ref{figD1}). Moreover, the `deterministic' estimation is always very close to the UBCM-induced, `annealed' one - a result that may be a consequence of the accurate description of the empirical network topology provided by the UBCM - evidently, much more accurate than those provided by the UBRGM and the LM.

Each solid line in Fig.~\ref{figD1} represents a normal distribution whose average value and variance coincide with the ones of the corresponding sample distribution: although the empirical and theoretical CDFs seem to be in (a very good) agreement, the Anderson-Darling test never rejects the normality hypothesis only for node 166 and does not reject the normality hypothesis in the case of the UBCM-induced distribution of values for node 168.\\

\begin{figure*}[ht!]
\centering
\includegraphics[width=0.48\textwidth]{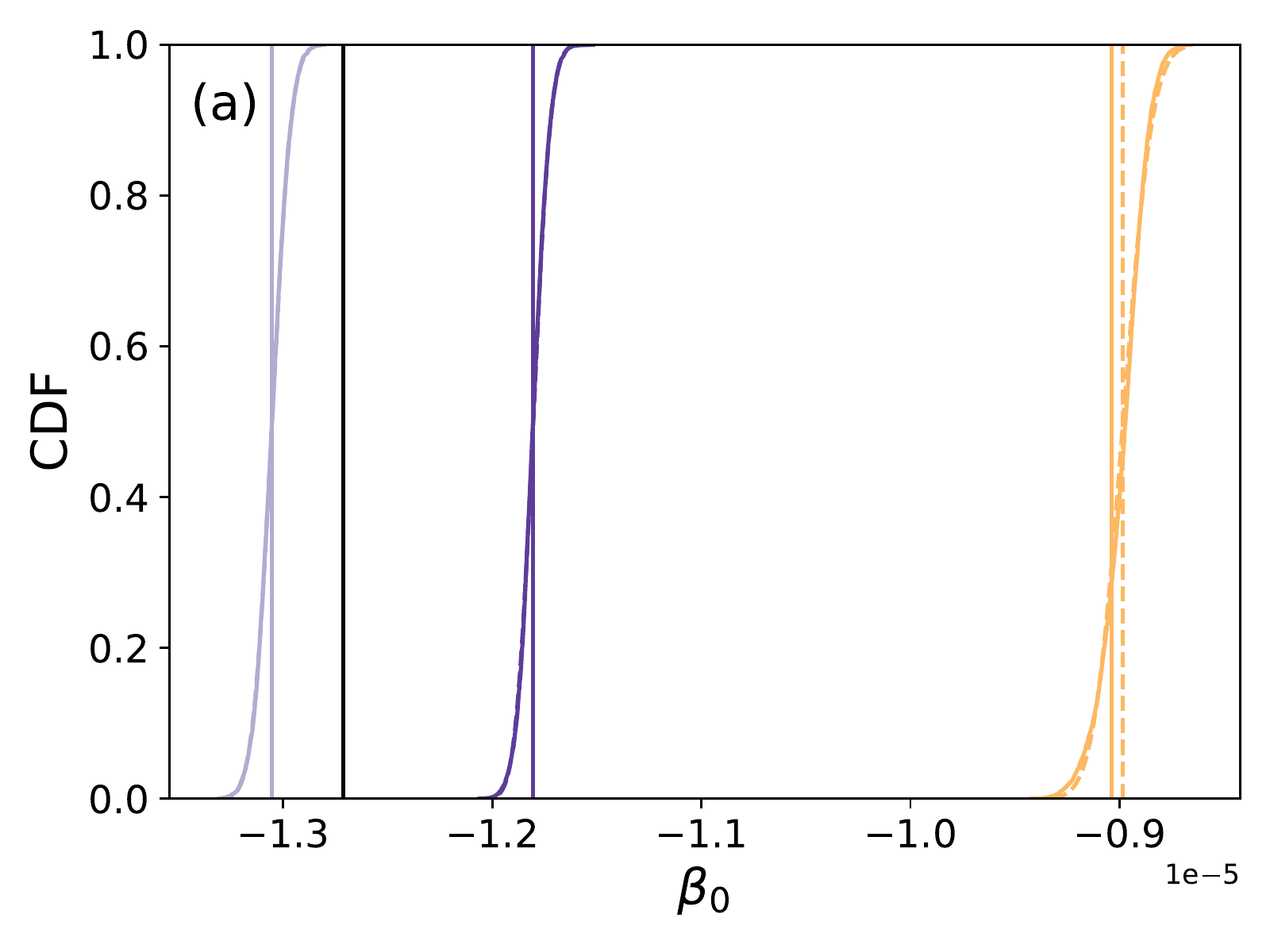}
\includegraphics[width=0.48\textwidth]{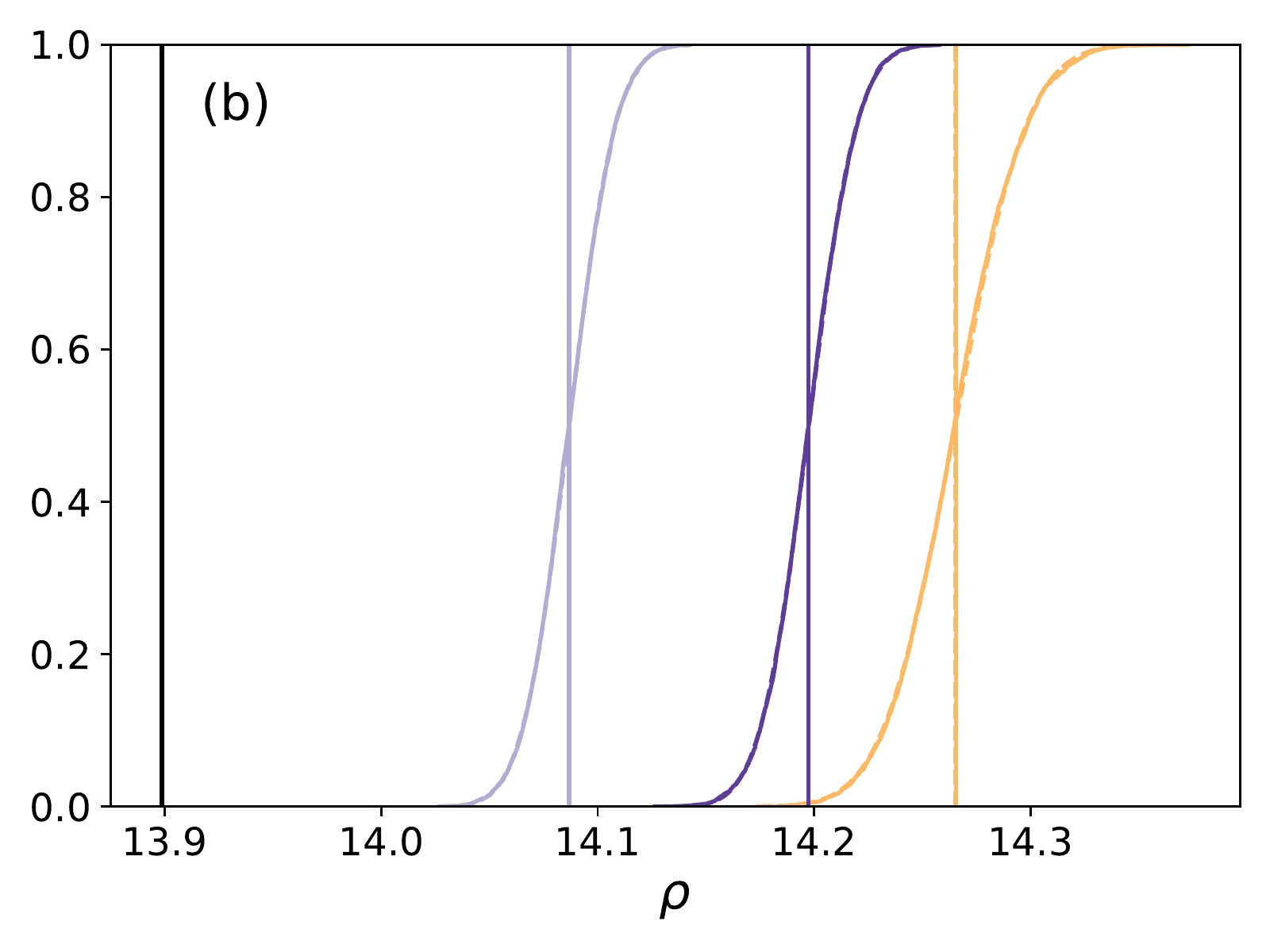}
\includegraphics[width=0.48\textwidth]{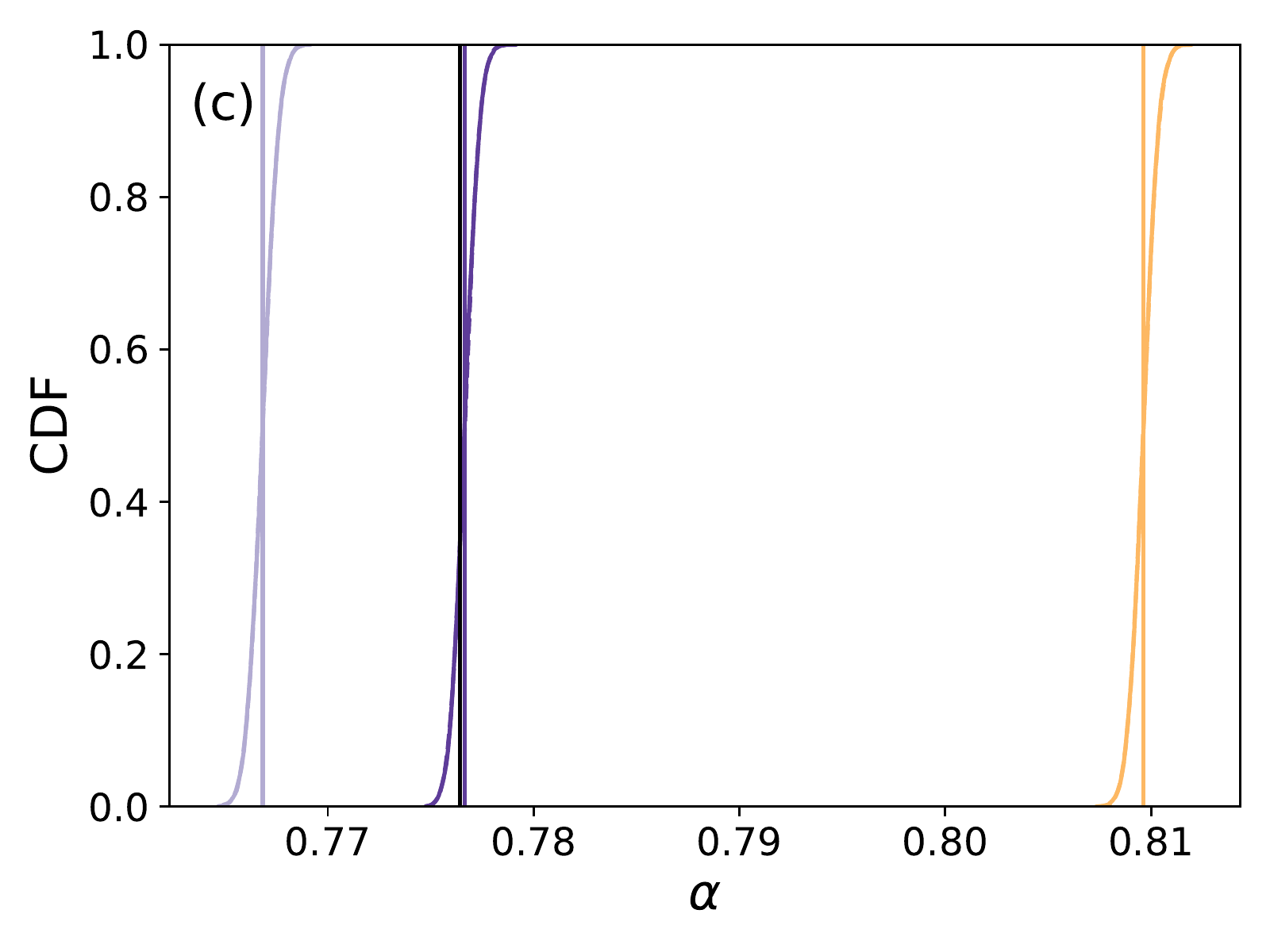}
\includegraphics[width=0.48\textwidth]{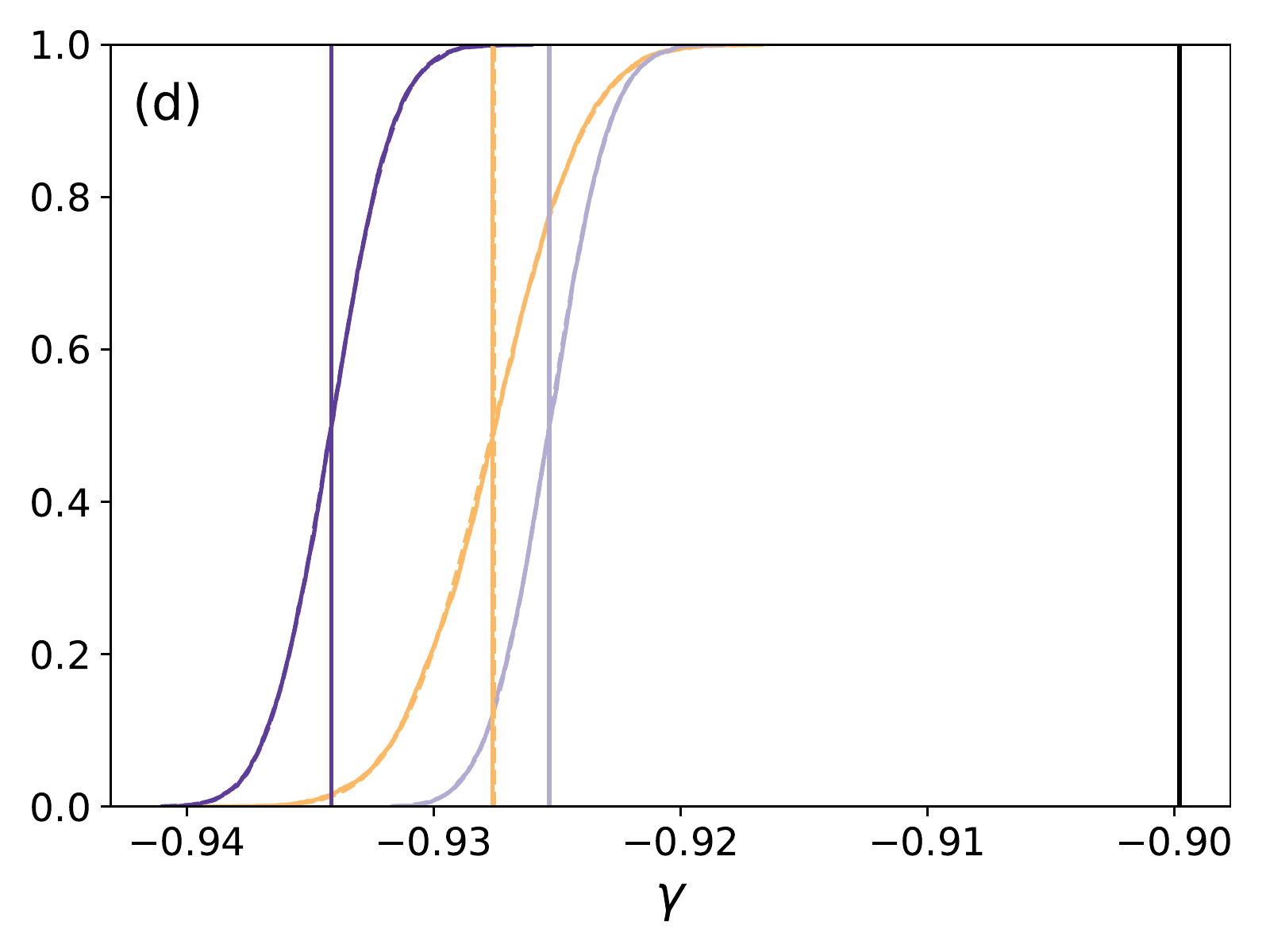}
\caption{Empirical CDFs for the parameters (a) $\beta_0$, (b) $\rho$, (c) $\alpha$ and (d) $\gamma$ entering the definition of the econometric version of the CEM, where the binary topology is either `deterministic' (black vertical line) or generated via the UBRGM (light orange or light grey), the UBCM (purple or dark grey) and the LM (light purple or grey). The deterministic approach leads to a single estimate, while the other approaches lead to either a single, `annealed' estimate (vertical, solid lines) or to a whole distribution of `quenched' estimates (constructed over an ensemble of 5.000 binary configurations; the corresponding average value is indicated by a vertical, dash-dotted line). The shapes of the `quenched', cumulative distributions induced by the three, binary recipes are very similar.}
\label{figD2}
\end{figure*}

\paragraph*{`Tensor' variant of the CEM.} Let us, now, leave $\beta_{ij}$ in its tensor form and constrain the set of weight-specific estimates $\hat{w}_{ij}$, $\forall\:i<j$. In this case, the three recipes lead to the following estimates

\begin{align}
\mathcal{L}_{\underline{\psi}}=\sum_{i<j}[-\beta_{ij}\hat{w}_{ij}+a_{ij}^*\ln\beta_{ij}]\quad&\Longrightarrow\quad\beta_{ij}=\frac{a_{ij}^*}{\hat{w}_{ij}}\\
\mathcal{G}_{\underline{\psi}}=\sum_{i<j}[-\beta_{ij}\hat{w}_{ij}+p_{ij}\ln\beta_{ij}]\quad&\Longrightarrow\quad\beta_{ij}=\frac{p_{ij}}{\hat{w}_{ij}}\\
\langle\beta_{ij}\rangle =\sum_{\mathbf{A}\in\mathbb{A}}P(\mathbf{A})\beta_{ij}(\mathbf{A})= \nonumber \\
= \sum_{\mathbf{A}\in\mathbb{A}}P(\mathbf{A})\frac{a_{ij}}{\hat{w}_{ij}}\quad&\Longrightarrow\quad\langle\beta_{ij}\rangle=\frac{p_{ij}}{\hat{w}_{ij}}
\end{align}

a result signalling large differences between the `deterministic' recipe, on the one hand, and the `quenched' and `annealed' recipes, on the other - that, instead, coincide. If, however, $\hat{w}_{ij}\equiv w_{ij}^*$, $\forall\:i<j$ then, for consistency, $p_{ij}\equiv a_{ij}^*$ and the three recipes coincide.

\paragraph*{`Econometric' variant.} As Figs. \ref{fig3} and \ref{figD2} show, the `deterministic' estimation is always quite different from the other, two ones - the only exception being represented by the parameter $\alpha$, under the UBCM-induced, binary recipe. Such a result should warn from employing the `deterministic' estimation recipe \emph{tout court} as ignoring the variety of structures that are compatible with a given probability distribution $P(\mathbf{A})$ will, in general, affect the estimation of the parameters of interest.

\end{document}